\documentclass[10pt,conference]{IEEEtran}

\ifCLASSINFOpdf
\else
\fi
\usepackage{textcomp}
\usepackage{amsmath}
\usepackage{fixltx2e}
\usepackage{balance}
\usepackage{mathptmx} % This is Times font
\usepackage{fancyhdr}
\usepackage[normalem]{ulem}
\usepackage[hyphens]{url}
\usepackage[draft]{hyperref}
\usepackage[normalem]{ulem}
\usepackage{booktabs}
\usepackage{multirow}
\usepackage[keeplastbox]{flushend}
\usepackage{xcolor}
\usepackage{gensymb}
\usepackage{listings}
\usepackage{enumerate}
\usepackage{indentfirst}
\usepackage{nicefrac}
\usepackage{tablefootnote}
\usepackage[para,online,flushleft]{threeparttable}
\usepackage{array}
\usepackage{cite}
\usepackage{tabularx}
\colorlet{light-gray}{gray!20}
\usepackage[bottom,hang,flushmargin,para]{footmisc}

\usepackage{graphicx}
\usepackage{caption}
\usepackage{subcaption}

\usepackage[latin9]{inputenc}

\usepackage{siunitx}
\usepackage{placeins}

\usepackage{amssymb}% http://ctan.org/pkg/amssymb
\usepackage{pifont}% http://ctan.org/pkg/pifont

\usepackage{romannum}

\usepackage{algorithm,algpseudocode}
\usepackage[font=small,labelfont=bf,
   justification=justified,
   format=plain]{caption}
\newcommand{\ignore}[1]{}

\makeatletter
\def\ps@IEEEtitlepagestyle{%
  \def\@oddfoot{978-1-5386-1233-0\/17\/\$31.00 \copyright2017 IEEE\hfill}%
  \def\@evenfoot{}%
}
\def\mycopyrightnotice{%
  {\footnotesize }% <--- Change here
  \gdef\mycopyrightnotice{}% just in case
}

\begin{document}
\title{Demystifying the Characteristics of 3D-Stacked Memories: A Case Study for Hybrid Memory Cube\vspace{-0.4in}}

\author{\IEEEauthorblockN{Ramyad Hadidi,
Bahar Asgari,
Burhan Ahmad Mudassar,\\
Saibal Mukhopadhyay,
Sudhakar Yalamanchili, and
Hyesoon Kim}
\IEEEauthorblockA{Email: \{rhadidi,bahar.asgari,burhan.mudassar,saibal.mukhopadhyay,sudha,hyesoon\}@gatech.edu
\\Georgia Institute of Technology}}

%Telephone: (800) 555--1212, Fax: (888) 555--1212}



\maketitle

%--------------------------%%%-----Abstract----%%%----------------------------------------%
\begin{abstract}
\noindent Three-dimensional (3D)-stacking technology, which enables 
the integration of DRAM and logic dies, offers high bandwidth and 
low energy consumption. 
This technology also empowers new memory designs for 
executing tasks not traditionally associated with memories. 
A practical 3D-stacked memory is Hybrid Memory Cube (HMC), 
which provides significant access bandwidth and low power 
consumption in a small area. 
Although several studies have taken advantage of the novel architecture of HMC, 
its characteristics in terms of 
latency and bandwidth 
or their correlation with temperature and power consumption have not been fully explored.
This paper is the first, to the best of our knowledge, 
to characterize the thermal behavior of HMC in 
a real environment using the AC-510 accelerator and to identify 
temperature as a new limitation for this state-of-the-art design space. 
Moreover, besides bandwidth studies, we deconstruct factors 
that contribute to latency
and reveal their sources for high- and low-load accesses.
The results of this paper demonstrates 
essential behaviors and performance bottlenecks  
for future explorations of packet-switched and 3D-stacked memories.
\vspace{-8pt}
\end{abstract}

% \fi
\IEEEpeerreviewmaketitle

\section{Introduction and Motivation}
\noindent
To date, the dominant architecture for computing systems has been
the processor-centric architecture, in which the processor and DIMM-based memories are 
separate units connected via the JEDEC protocol~\cite{jedec-ddr4}. 
This architecture inherently enables the use of a large centralized memory
accessed by multiple processors. 
However, the quest for increasing memory bandwidth has led to the development of 3D-DRAM
packages and the resulting emergence of on-package, high-bandwidth 3D
memory tightly integrated with the processor, two examples of which are
High Bandwidth Memory (HBM)~\cite{lee:kim14} and  
Hybrid Memory Cube (HMC)~\cite{jed:kee12}.
While HBM provides high bandwidth on top of the traditional JEDEC-based communication, 
HMC utilizes modern packet-based interfaces and relies on high degree of
internal concurrency. 
Further, to eliminate costly data
movement between the processor and memory subsystems, researchers are
exploring the integration of compute units within such high-bandwidth
memories (e.g., the integration of CMOS logic and DRAM dies within a
3D stack). 
This paper reports on experimental characterizations of
HMC to provide an understanding of sources of high performance and
bottlenecks; and therefore,	 implications for effective
application design for both traditional and processor-in-memory (PIM)
configurations.

The novel structure of HMC has motivated researchers to develop
architectures based on concepts introduced in HMC~\cite{pug:jes14,
nai:had17, zha:jay14, kim:kim13, hsi:ebr16}.  
Although many researchers have studied the
3D-stacked layered integration of computation and memory based on
simulation, experimental studies are 
few~\cite{sch:fro2016, ibr:fat2016, gok:llo15}.
In particular, to the best
of our knowledge, no experimental work has sought to characterize
the relationship between temperature, power consumption, bandwidth, and 
latency\footnotemark\,of these memories. 
%
%For example, high bandwidth operation may increase power consumption 
%and thus increased temperature. 
% >>>>>>>>> for space
%Furthermore, like DRAM, HMC dynamically
%changes the refresh rate of DRAM banks based on temperature conditions
%and detected error rates~\cite{hybrid2013hybrid1} further increasing
%power consumption. 
%
Importantly, in PIM configurations, 
a sustained operation can eventually lead to failure by exceeding 
the operational temperature of HMC, assumed to be 
$85$\textdegree C~\cite{zhu:yux:2016, yas:nuw14}. 
Furthermore, the physical organization of 3D-stacked memories introduces new
structural abstractions with corresponding new latency and bandwidth
hierarchies. 
\footnotetext{This paper uses the term latency and round-trip time
interchangeably.}
Therefore, maximizing
bandwidth utilization relies on data mappings and associated reference
patterns that are well matched to the concurrency of the internal
organization of HMC. 
Our analysis examines the implications of
this internal organization on the mapping of data structures in applications.

\begin{figure}[b]
\centering
\vspace{-0.2in}
\includegraphics[width=0.95\linewidth]{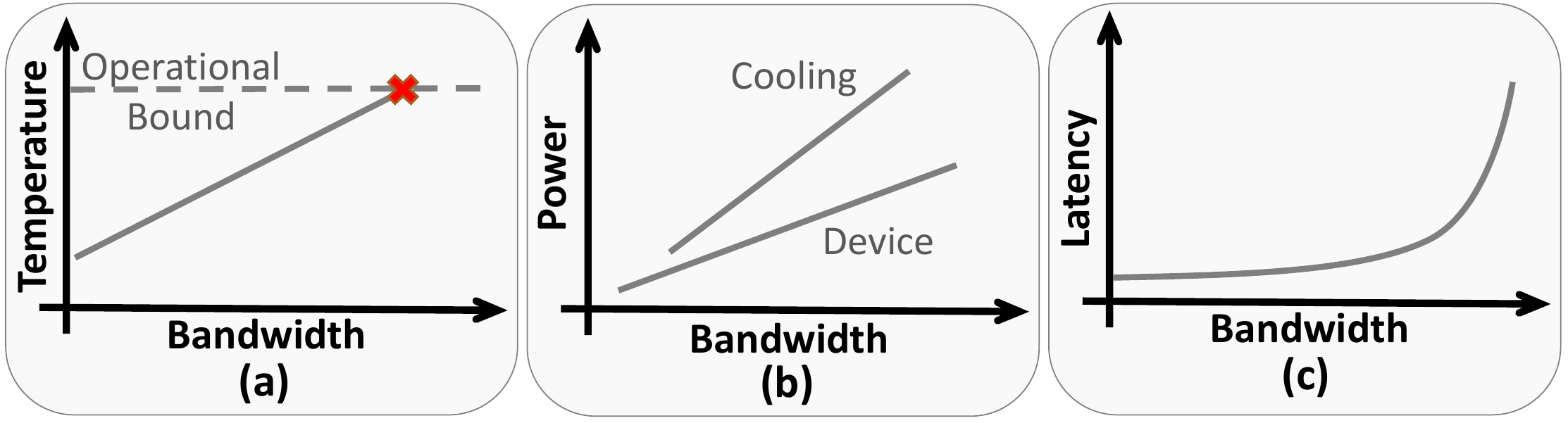}
\vspace{-4pt}
\captionsetup{singlelinecheck=on,aboveskip=5pt, belowskip=-4pt}
\caption{Conceptual graphs showing relationships between the
temperature, power consumption, latency, and bandwidth of HMC.}
\label{fig:moti}
\vspace{-0.0in}
\end{figure}
Characterizations in this paper, as shown in Figure~\ref{fig:moti},
seek to explore and quantify the conceptual
relationships between bandwidth, temperature, power consumption, and
latency.
Figure~\ref{fig:moti}a depicts the first relationship: 
as bandwidth increases, temperature increases that can eventually exceed the operational range.
To mitigate high temperatures, systems must use costlier cooling
solutions.\footnotemark 
\footnotetext{For instance, AMD Radeon R9 Fury
X GPU, which integrates an HBM and a GPU on an interposer, requires
a liquid cooling solution~\cite{macri2015amd}.}  
\,Moreover, higher temperatures trigger mechanisms such as
frequent refresh~\cite{hybrid2013hybrid1}, 
which also increases power consumption.
Therefore, as Figure~\ref{fig:moti}b depicts, as bandwidth increases,
the power consumption of a device and its required cooling power
rises.  
In addition, to determine the benefits of HMC in a system, an
integral asset is the relationship between the bandwidth and latency (Figure~\ref{fig:moti}c).  
Since we use a prototype
infrastructure, we deconstruct elements that contribute to
latency under low- and high-load utilizations with various bandwidth profiles.  
We believe that
understanding the impact of such elements provides insights about the
full-system impact of the HMC in particular, and 3D-stacked memories in
systems in general.

To this end, in this paper, we use a system with a controllable cooling system and
an AC-510~\cite{ac510} accelerator board,
which includes an HMC\,1.1 (\emph{Gen2})~\cite{hybrid2013hybrid1} and an FPGA.
On the FPGA, we systematically generate synthetic workloads using customized Verilog implementations based on \emph{GUPS} (giga updates per second) such as various combinations of high-load, low-load, random, and linear access patterns, which are building blocks of real applications.    
We analyze the bandwidth, latency, and power consumption of HMC while using a thermal
camera to record its temperature.
This paper contributes the following:
(\romannum{1}) This is the first study, to the best of our knowledge, that
  measures the effects of increased bandwidth on temperature and power altogether in a real 3D-stacked memory system and explores the sensitivities to, and consequences of, thermal limits;
(\romannum{2}) it exposes the interactions among
  bandwidth utilization, power consumption, and access
  patterns in a real 3D-stacked memory system;
(\romannum{3}) it explores contributing factors to latency and
  yields insights into why, when, and how HMC benefits applications; and
(\romannum{4})  it presents an analysis of latencies in HMC based on 
  the impact of packet-switched interfaces, and 
  the relationship to the structural organization (i.e., vaults, banks, and quadrants).

\section{Hybrid Memory Cube}
\label{hmc}
\noindent
This section introduces the architectural organization, communication
interface and protocol, and address space mapping of the HMC. We focus
on the HMC\,1.1 specification~\cite{hybrid2013hybrid1}
for which the hardware is currently available and used in the experiments
described in this paper. We also briefly compare
the specification of HMC\,2.0~\cite{hybrid2014hybrid}, whose hardware is
yet to be available, HMC\,1.1 (\emph{Gen2}), and HMC\,1.0
(\emph{Gen1})~\cite{hybrid2013hybrid}.

%----------------------------------------------------%
\subsection{HMC Structure}
\label{sec:hmc-struc}
\begin{figure}[b]
\centering
\vspace{-0.24in}
\includegraphics[width=0.82\linewidth]{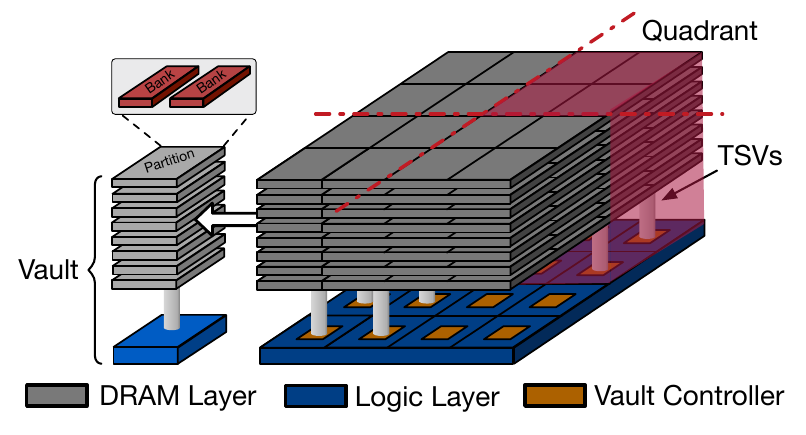}
\vspace{1pt}
\captionsetup{singlelinecheck=on,aboveskip=-1pt,belowskip=1pt}
\caption{4\,GB HMC~1.1 internal structure.}
\label{fig:hmc-struc}
\vspace{-0.10in}
\end{figure}

\noindent
HMC is a 3D-stacked memory composed of a logic die upon which multiple DRAM
layers are stacked.
The dies are vertically connected by \emph{through-silicon-vias
(TSVs)}, which provide higher internal bandwidth, lower latency, and lower
communication energy consumption within a cube than comparable 2D
structures~\cite{zha:jay14,jed:kee12,paw11}.
Figure~\ref{fig:hmc-struc} shows the HMC\,1.1 structure, composed of
one logic layer and eight DRAM layers, each divided into 16 partitions.
Accordingly, the 3D stack is divided into 16 vertical
\emph{vaults}~\cite{hybrid2013hybrid},
each with its own memory controller in the logic layer, connected to DRAM
partitions by 32 data TSVs\footnotemark\,reside above it~\cite{jed:kee12}.
\footnotetext{
Another reference~\cite{hybrid2013hybrid1} mentions that internally
within each vault the granularity 
of the DRAM data bus is 32\,B.}
Therefore, in HMC\,1.1, each memory controller is responsible for eight partitions
of DRAM layers.
As shown in the figure, a collection of four vaults is called a quadrant,
whose vaults share a single external link.\footnotemark\,
\footnotetext{As shown in Table~\ref{tab:hmc-struc},
each quadrants contains 8 vaults in HMC\,2.0.}
Each DRAM layer in a Gen1 device (HMC\,1.0) is a $68\,mm^2$
1\,Gb die manufactured
in 50\,nm technology~\cite{jed:kee12}.
Each Gen1 device has four layers of DRAM and the size of a Gen1 device is
$(1\,\text{Gb} \times 4) \div 8\,\text{bits/B} = 512\,\text{MB}$.
Further, each layer is composed of 16 partitions, so the size of each DRAM
partition is
$1\,\text{Gb} \div 16 = 64\,\text{Mb} = 8\,\text{MB}$.
In addition, in Gen1 devices, each DRAM partition has two
independent DRAM banks~\cite{jed:kee12}, each
$8\,\text{MB} \div 2 = 4\,\text{MB}$.
In Gen2 devices, the number and size of DRAM layers
increase from four to eight and 1\,Gb to 4\,Gb, respectively, with the same number of partitions.
Thus, each partition is 32\,MB, and each bank is
16\,MB~\cite{hybrid2013hybrid1}.
Therefore, the number of banks in each Gen2 device is equal to
{\small
\vspace{-3pt}
\begin{equation}
\begin{split}
    \text{\#Banks}_{HMC\,1.1} =&
	8\,\text{layers} \times 16\,\nicefrac{\text{partitions}}{\text{layer}}
	\times 2\,\nicefrac{\text{banks}}{\text{partition}} \\
	=& 256\,\text{banks.}
	\label{eq:banks}
\end{split}
\end{equation}
}
For HMC\,2.0, the number of vaults increases to 32, and
DRAM layers are denser.
Table~\ref{tab:hmc-struc} summarizes the structural properties of each
generation of HMC devices.
%
%%%%%%
%%%%%%
%%%%%%
\renewcommand{\arraystretch}{0.8}
\begin{center}
\vspace{-16pt}
\small
\begin{threeparttable}[h]
\captionsetup{singlelinecheck=on,aboveskip=1.5pt}
\caption{Properties of HMC
         Versions~\cite{hybrid2013hybrid1,hybrid2013hybrid,hybrid2014hybrid,
          jed:kee12}.}
	\begin{tabular}{r| c | c | c}
		%\cmidrule[0.08em]{2-4}
		\toprule
		 &  \textbf{HMC\,1.0} & \textbf{HMC\,1.1}
		 & \multirow{2}{*}{\textbf{HMC\,2.0}\textdagger}\\
		 &  \textbf{(Gen1)}\tnote{\textdagger}
		 &  \textbf{(Gen2)}\tnote{\textdagger} & \\
		\midrule
		%\cmidrule{2-4}
		Size & 0.5\,GB & 2/4\,GB & 4/8\,GB \\
		\#~DRAM Layers & 4 & 4/8 & 4/8 \\
		DRAM Layer Size & 1\,Gb & 4\,Gb & 4/8\,Gb \\
		\#~Quadrants & 4 & 4 & 4 \\
		\#~Vaults & 16 & 16 & 32 \\
		Vault/Quadrant & 4 & 4 & 8\\
		\#~Banks & 128 & 128/256 & 256/512 \\
		\#~Banks/Vault & 8 & 8/16 & 16/32 \\
		Bank Size & 4\,MB & 16\,MB & 16\,MB \\
		Partition Size & 8\,MB & 32\,MB & 32\,MB \\
		\bottomrule
	\end{tabular}
		\begin{tablenotes}
    	\item[\textdagger]
    	 \footnotesize{Reported values are for the four-link configuration.
    	 While HMC\,2.0 only supports the four-link configuration, HMC\,1.x 
    	 supports both four- and eight-link configurations.}
    	\end{tablenotes}
	\vspace{-0.00in}
	\label{tab:hmc-struc}
\end{threeparttable}
\end{center}
\renewcommand{\arraystretch}{1}
%%%%%%
%%%%%%
%%%%%%

%----------------------------------------------------%
\subsection{HMC Communication Protocol}
\label{sec:hmc-inter}

\noindent
The HMC interface utilizes a packet-based communication protocol
implemented with high speed serialization/deserialization
(\emph{SerDes}) circuits.  Hence, implementations achieve
higher raw link bandwidth than achievable with synchronous bus-based
interfaces implementing the traditional JEDEC protocol.  Packets are
partitioned into 16-byte elements, called \emph{flits}.  Supported
packet sizes for data payloads range from one flit (16\,B) to eight
flits (128\,B).  Each packet also carries an eight-byte header and
an eight-byte tail; therefore, each request/response has an overhead
of one flit~\cite{sch:fro2016}.  The header and tail
ensure packet integrity and proper flow
control~\cite{hybrid2013hybrid1}.  Table~\ref{tab:hmc-rdwrsize} shows
each HMC transaction size in flits.
\renewcommand{\arraystretch}{0.95}
\begin{table}[h]
    %\vspace{-0.1in}
	\footnotesize
	\centering
	\vspace{-12pt}
	\captionsetup{singlelinecheck=on,aboveskip=1pt}
	\caption{HMC read/write request/response sizes~\cite{hybrid2013hybrid1}.}
	\begin{tabular}{r| c c | c c}
		\toprule
		\multirow{2}{*}{\textbf{Type}} &
		\multicolumn{2}{c|}{\textbf{Read}} &
		\multicolumn{2}{c}{\textbf{Write}} \\
		& Request & Response & Request & Response \\
		\midrule
		Data Size & Empty & 1$\sim$8\,Flits & 1$\sim$8\,Flits & Empty\\
		Overhead & 1\,Flit & 1\,Flit & 1\,Flit & 1\,Flit   \\
		\midrule
		Total Size & 1\,Flit & 2$\sim$9\,Flits & 2$\sim$9\,Flits & 1\,Flit \\
		\bottomrule
	\end{tabular}
	\vspace{-7pt}
	\label{tab:hmc-rdwrsize}
\end{table}
\renewcommand{\arraystretch}{1}
To connect to other HMCs or hosts, an HMC uses two or four external
links.  Each independent link is connected to a quadrant that is
internally connected to other quadrants, which routes
packets to their corresponding vaults.
As a result, an access to a local vault in a quadrant incurs lower
latency than an access to a vault in another
quadrant~\cite{hybrid2013hybrid1}.  Furthermore, each external link is
a 16- (full-width) or eight-lane (half-width) connection that supports
full-duplex serialized communication across each bit lane with configurable speeds of 10,
12.5, or 15\,Gbps.  For instance, a two-link half-width HMC device
whose link operates at 15\,Gbps has a maximum bandwidth of
\vspace{-2pt}
\begin{equation}
\small
\label{eq:bw}
\begin{split}
  \text{BW}_\text{peak} =& 2\,\text{link} \times
              8\,\nicefrac{\text{lanes}}{\text{link}} \times
             15\,\text{Gbps} \times 2\,\small{\text{full duplex}}
             \\=& 480\,\text{Gbps} = 60\,\text{GB/s.}
\end{split}
\end{equation}

%----------------------------------------------------%
\subsection{HMC Memory Addressing}
\label{sec:hmc-add}

\noindent
This section presents the internal address mapping used in HMC and
explores some of its implications and opportunities for optimization.
In HMC, DRAM operation follows a closed-page policy~\cite{jed:kee12};
therefore, on completion of a memory reference, the sense amplifiers
are precharged and the DRAM row is closed.  The page size (i.e., row size)
in HMC is 256\,B~\cite{jed:kee12}, smaller than that
in DDR4~\cite{jedec-ddr4} which varies from 512 to 2,048\,B.
Since HMC has several banks, keeping DRAM rows open incures  
high power consumption, so HMC employs the closed-page policy.
This policy and the small page size
reduce the power consumption for accesses with low temporal locality
and high miss ratio.
Moreover, internally within each vault, the granularity of the DRAM data
bus is 32\,B~\cite{hybrid2013hybrid1}.
Thus, as the specification points out,
starting or ending a request on a 16-byte boundary uses the
DRAM bus inefficiently . 
As a result, an application that targets bandwidth optimization
should issue requests on 32-byte boundaries.

\begin{figure}[t]
\centering
\vspace{-0.1in}
\includegraphics[width=0.82\linewidth]{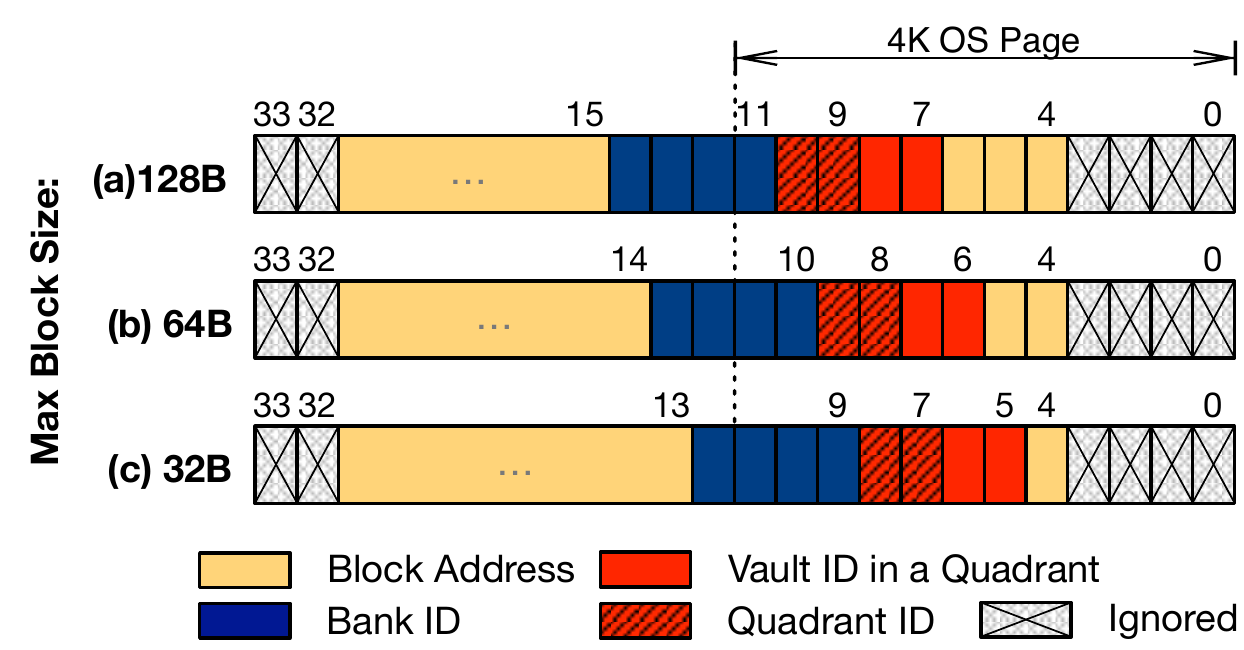}
\vspace{4pt}
\captionsetup{singlelinecheck=on,aboveskip=-0pt,belowskip=1pt}
\caption{Aaddress mapping of 4\,GB HMC\,1.1 with various maximum block size of (a) 128\,B, (b) 64\,B, and (c) 32\,B~\cite{hybrid2013hybrid1}.}
\label{fig:hmc-address}
\vspace{-0.25in}
\end{figure}

The request header of HMC contains a 34-bit address
field (16\,GB addressable memory). However, because current hardware does not 
support a 16\,GB capacity, the two high-order address bits are
ignored.
HMC employs a \emph{low-order-interleaving} policy for mapping the
memory blocks across vaults and banks as illustrated in
Figure~\ref{fig:hmc-address}. 
The block size is 16\,B, so low-order four bits are ignored. 
Next address bits define the maximum 
block size in accessing the HMC. Figure~\ref{fig:hmc-address}a illustrates mapping for 
the default maximum block size of 128\,B.\footnotemark\,
The next four address bits are used to identify
a vault followed by four bits to identify a bank within the selected
vault. Thus, sequential blocks are first distributed across various vaults and
then banks. According to the specification, the user may fine-tune the address mapping
scheme by changing bit positions used for vault and bank mapping.
This paper studies the default address mapping of
HMC.\footnotemark[\value{footnote}] 
\footnotetext{
The maximum block size is controlled with the Address Mapping Mode Register to
be 16, 32, 64, or 128\,B. 
The default mapping is defined by setting Address Mapping Mode Register 
to \texttt{0x2}, or 128\,B max block size.
}

%------------------------------------------------

%>>>>>>>>>>>> for exp setup section
%
\begin{figure*}[]
  \vspace{-0.2in}
  \begin{tabular}{c | c}
  \begin{subfigure}{0.70\columnwidth}\centering
  \includegraphics[width=\textwidth]{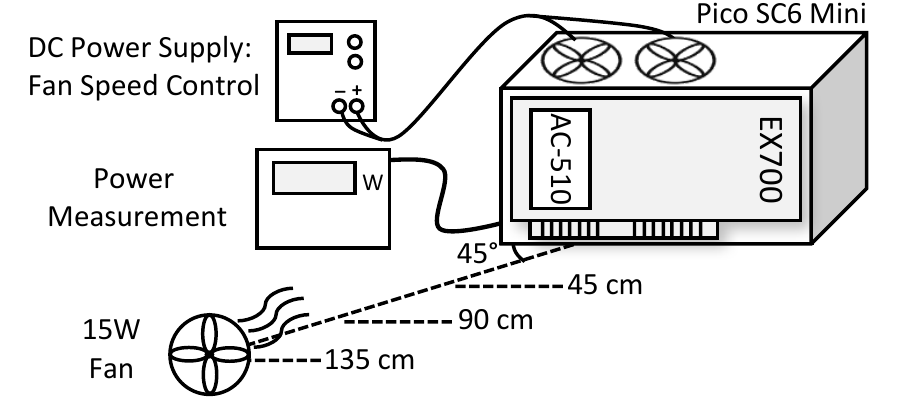}
  \captionsetup{singlelinecheck=on,aboveskip=-9pt,belowskip=-5pt,font=small}
  \caption{}
  \label{fig:meth-infs}
  \end{subfigure}
  &
  \begin{subfigure}{1.30\columnwidth}\centering
  \includegraphics[width=\textwidth]{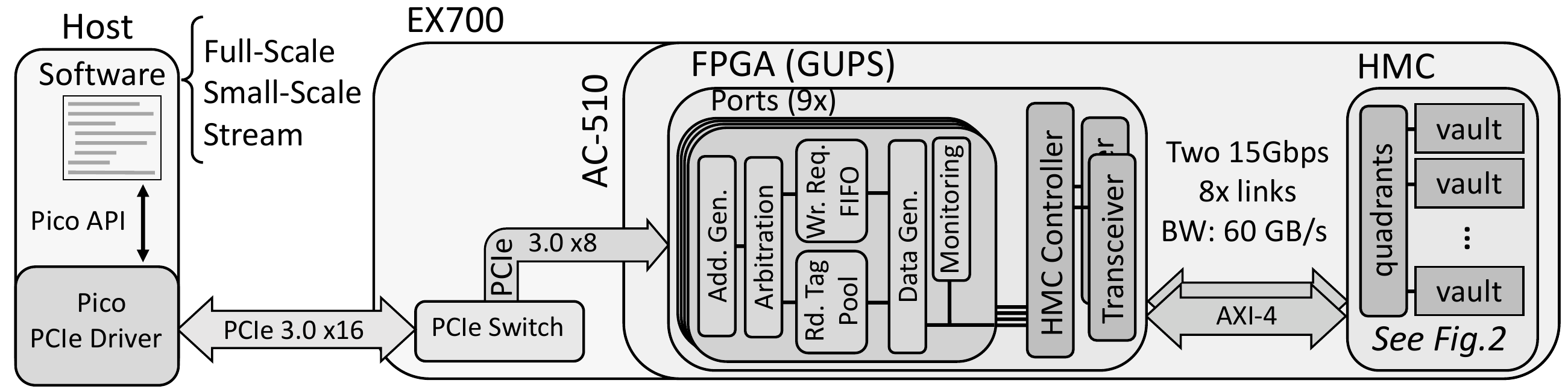}
  \captionsetup{singlelinecheck=on,aboveskip=-9pt,belowskip=-5pt,font=small}
  \caption{}
  \label{fig:sys-gups}
  \end{subfigure}
  \\
  \end{tabular}
  \captionsetup{singlelinecheck=on,aboveskip=5pt}
  \caption{Infrastructure overview: (a) Cooling environment configuration,
  (b) firmware and software overview.}
  \vspace{-0.2in}
  \label{fig:infs}
  \vspace{-0.1in}
\end{figure*}
In a higher level of abstraction, 
a 4\,KB OS page is allocated in two banks across all vaults
(HMC\,1.1),\footnote{Note that two banks is when the max block size is set to 128\,B.
We can increase BLP in accessing a single page
by reducing the max block size (Figure~\ref{fig:hmc-address}).}
promoting memory level parallelism for
reading/writing the contents of a page.
When we access multiple pages allocated serially in the physical
address space, as the number of concurrent accesses to these pages increases,
bank-level parallelism (BLP) increases as well.
For instance, in a 4\,GB HMC\,1.1 with 16 banks per vault,
because a single page occupies two banks in a vault, up to eight
allocated pages reside in a vault.
Therefore, for 16 vaults, the total number of pages we can access 
while increasing BLP is
$16\,\text{vaults} \times 8\,\nicefrac{\text{\#pages}}{\text{vault}} = 128$.
In fact, combined with BLP that each page utilizes, the
low-order-interleaved address mapping enables maximum BLP for
sequential page accesses.
To conclude, 3D DRAM exemplified by HMC presents 
new address space abstractions
and associated performance implications, the understanding 
of which is necessary for 
optimizing compilers and data layout
algorithms for maximizing performance potential. 
%
%In a scenario that multiple pages are accessed randomly, the frequency, size, and
%coverage (in terms of number of banks, vaults, and rows) of accesses
%determine the performance and bandwidth of an application.
%
In a scenario that an application randomly accesses multiple pages, 
the frequency, size, and
coverage (in terms of number of banks, vaults, and rows) of accesses
determine performance.
In this paper, we explore HMC design space to understand its 
full-scale impact on applications.

\section{Experimental Setup}

\label{sec:meth}
\noindent This section introduces the infrastructure,
cooling equipment, and firmware, all of which are shown in Figure~\ref{fig:infs}.

%-----------------------------------------------------%
\subsection{Infrastructure Overview}
\label{sec:meth-inf}
%
%\noindent
\hspace{-0.02in} \textbf{\emph{Hardware:}}
Our infrastructure, Pico SC-6 Mini~\cite{SC6Mini}, contains an EX700~\cite{ex700}
backplane, a PCIe\,3.0\,x16 board with 32\,GB/s bandwidth.
EX700 integrates a PCIe switch that routes host communication to
up to six AC-510~\cite{ac510} accelerator modules via PCIe\,3.0\,x8 buses.
Each module has a Kintex UltraScale Xilinx
FPGA\footnotemark\,
connected to a 4\,GB HMC Gen2 (the same as Figure~\ref{fig:hmc-struc}) 
with \emph{two} half-width (8 lanes) links operating
at 15\,Gbps, so the bi-directional peak bandwidth is 60\,GB/s, as
Equation~\ref{eq:bw}.
%This paper studies an AC-510 accelerator.
%
\footnotetext{Part\#: xcku060-ffva1156-2-e}
%
%

%-----------------------------------------------------%
\par
\begin{figure}[t]
\centering
\vspace{-0.00in}
\includegraphics[width=0.9\linewidth]{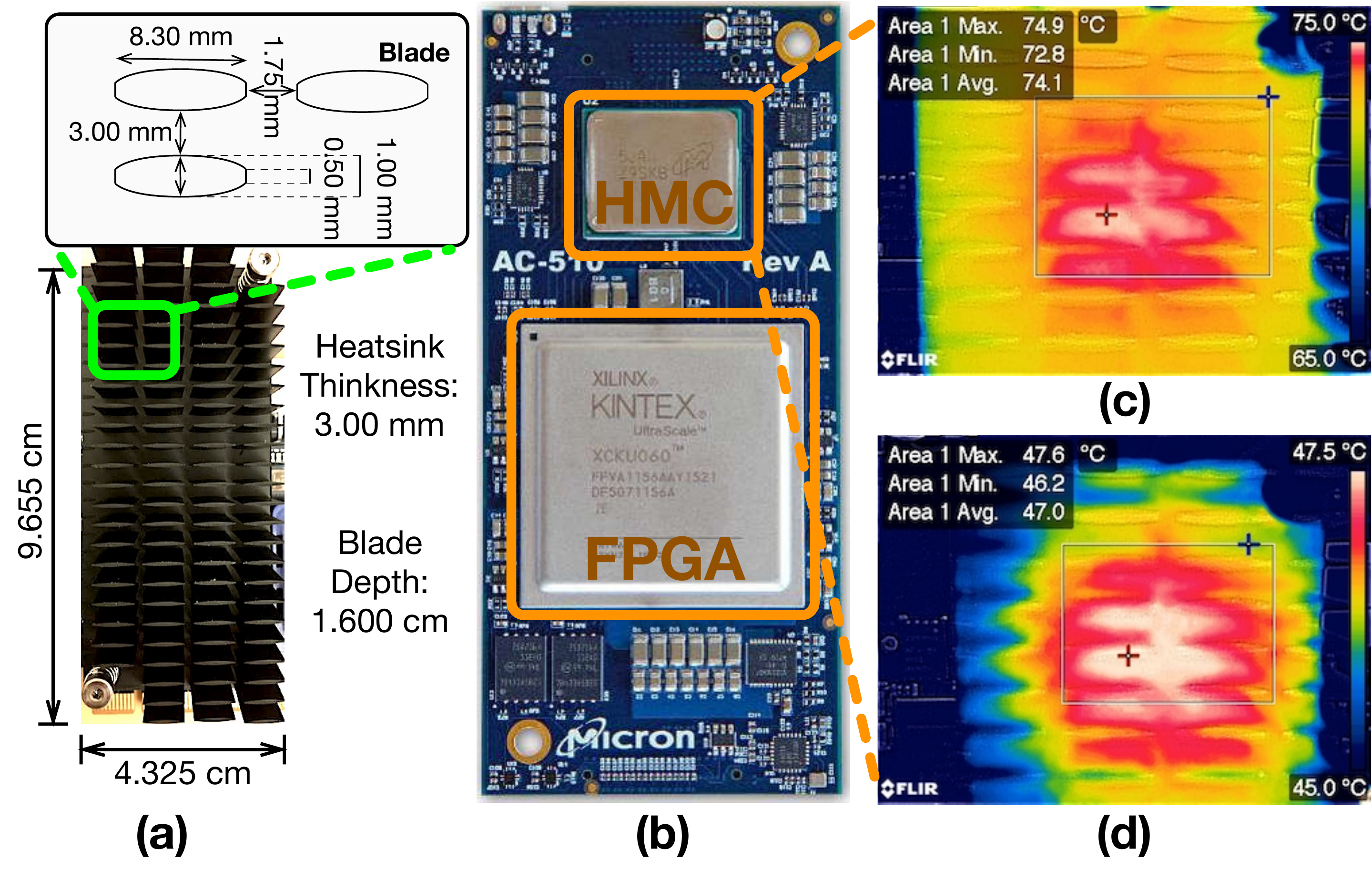}
\captionsetup{singlelinecheck=on,aboveskip=3pt}
\caption{AC-510 accelerator with (a) and without the heatsink (b), and
two images of the HMC with various temperature, taken by the thermal camera showing heatsink surface temperature (c,d).\protect\footnotemark}
\vspace{-0.3in}
\label{fig:meth-ac510}
\end{figure}

\textbf{\emph{Thermal Measurements:}}
To cool both the FPGA and HMC, an active heatsink 
equivalent to a low-end active cooling solution with
{\small$\sim$}\$30 cost~\cite{zhu:yux:2016} is attached on top of 
the accelerator module. 
Figure~\ref{fig:meth-ac510}a and \ref{fig:meth-ac510}b 
display the accelerator module with and without the heatsink, respectively.
For thermal evaluations, we run each experiment for 200 seconds,\footnotemark\,and
we measure the \emph{heatsink surface temperature}
of HMC using a thermal camera (Flir One~\cite{flir-camera}).
\footnotetext{For all experiments, after 200 seconds, temperature is stable.}
Since thermal resistance of a transistor chip is
smaller than that of the external heatsink,
temperature of the heatsink surface is 5--10 degrees Celsius lower than that
of the in-package junction~\cite{eric:road97}.
The FPGA and HMC share the heatsink, but HMC is
distinguishable in thermal images,
and it has a higher temperature than its surroundings do,
as seen in the thermal images of HMC at two temperatures in
Figure~\ref{fig:meth-ac510}c and \ref{fig:meth-ac510}d.  
\footnotetext{To display the heat island created by
HMC, we modify the heat scale, also shown at the right of each image.}
In other words, the shared heatsink does not prevent us from observing the effect
of temperature variation because
(\romannum{1}) HMC creates a heat island, which indicates it causes temperature
increments,
(\romannum{2}) FPGA performs the same tasks during experiments, so its heat dissipation is constant, and
(\romannum{3}) in real systems, one component in the box is always affected by the heat profile of other components.
In near-processor architectures, the temperature of an HMC is always
affected by the temperature of a computational component such as a
GPU, a CPU, or an FPGA.   
The thermal coupling between the FPGA and HMC is a factor of the integration substrate.   
Our measurements represent the PCB coupling, a much weaker coupling than integration of 
compute and memory in interposer or 3D.
Thus, the configuration of our experiments is a realistic representation of a future
design. 

\renewcommand{\arraystretch}{0.8}
\begin{table}[b]
	\footnotesize
	\centering
	\vspace{-19pt}
	\captionsetup{singlelinecheck=on,aboveskip=1pt}
	\caption{Experiment cooling configurations.}
	\begin{tabular}{c | c | c | c | c}
		\toprule
        Configuration & \multicolumn{2}{c|}{DC Power Supply:} & 15\,W Fan & Average HMC \\
        Name    & Voltage & Current & Distance  & Idle Temperature \\
		\midrule
		\texttt{Cfg1} & 12\,V         & 0.36\,A                 & 45\,cm  & $43.1\degree$\,C\\
		\texttt{Cfg2} & 10\,V         & 0.29\,A                 & 90\,cm  & $51.7\degree$\,C\\
		\texttt{Cfg3} & 6.5\,V        & 0.14\,A                 & 90\,cm  & $62.3\degree$\,C\\
		\texttt{Cfg4} & 6.0\,V        & 0.13\,A                 & 135\,cm & $71.6\degree$\,C\\
		\bottomrule
	\end{tabular}
	\label{tab:cooling}
	\vspace{-0.0in}
\end{table} \renewcommand{\arraystretch}{1}
%-----------------------------------------------------%
\par
\textbf{\emph{Cooling Environment:}}
The PCIe backplane resides in a case and is air cooled via 
two fans attached to the top of the case\footnotemark\,(total measured power of 4.5\,W with 12\,V).
\footnotetext{These fans are for cooling \emph{only} the backplane and not the machine.}
To mimic the thermal impact of PIM techniques,
which is a potential application for 3D-stacked memories, we create various
thermal environments (see Table~\ref{tab:cooling}) by tuning the speed of the fans
with a DC power supply, and placing a commodity fan (Vornado Flippi V8)
with a measured power of 15\,W and
at an angle of $45\degree$ at three distances (45, 90, and 135\,cm).
Figure~\ref{fig:meth-infs} shows a diagram of our
cooling environment.

%-----------------------------------------------------%
\par
\textbf{\emph{Power Measurements:}}
During the experiments, with a power analyzer, we monitor the
power consumption of the machine, which includes the power of the
FPGA, HMC, and PCIe switch on the backplane, and report the
average of the power usage.
Cooling fans are connected to another power source.
Since during experiments the host and FPGA do not communicate,
the power consumption of the PCIe switch is negligible.
Thus, power consumption above the idle power of the machine,
100\,W, is counted towards the activity of the HMC and
the FPGA.
In our infrastructure, we cannot decouple the power consumption
of these two components since they share the same board and power supply.
However, since the FPGA performs same task in all the experiments, 
variation in measured power is attributed to the HMC.

%-----------------------------------------------------%
\subsection{Firmware and Software Overview}
\label{sec:meth-firm}
\noindent
We use various firmware (i.e., digital design on the FPGA)
and software combinations to perform experiments.
To measure the properties of accesses
to the HMC with various address patterns,
we use the Pico API~\cite{pico-api} and a Verilog implementation of \emph{GUPS},
which measures how frequently we can generate
requests to access memory locations (Figure~\ref{fig:sys-gups}).
The FPGA uses Micron's controller~\cite{hmc-controller} to generate packets for the
multi-port AXI-4 interface between the FPGA and HMC.
The Pico API initializes the hardware and provides an environment with the host OS
to communicate with the FPGA on each accelerator module.
Although the Pico API provides an interface to access HMC through the FPGA,
since its read and write operations are bundled with software, 
a pure software solution to measure the bandwidth 
lacks sufficient speed.
Therefore, to generate and
send various mixes of requests, we use GUPS,
which utilizes the full potential of available HMC bandwidth.
As the frequency of the FPGA is low (187.5\,MHz),
to saturate the available bandwidth, GUPS uses nine copies of the same module,
or \emph{ports} to generate requests.
Each port includes a configurable address generator, a monitoring unit to measure
read latencies, a read tag pool with the depth of 64, and an arbitration unit
to generate the type of a request (read or write).
Each port is configurable to send
\emph{read only} (\texttt{ro}), \emph{write only} (\texttt{wo}),
or \emph{read-modify-write} (\texttt{rw}) requests
for \emph{random}, or \emph{linear} mode of addressing.
Furthermore, requests can be mapped to a specific part of the 
HMC by forcing some bits of the address to zero/one by using 
address \emph{mask}/\emph{anti-mask} registers.
We use the following three implementations:
\par
\textbf{\emph{Full-scale GUPS:}}
To measure bandwidth, temperature, and high-load latency of various
access patterns, we use full-scale GUPS, which utilizes all nine ports.
For such experiments, first, we activate the ports, set the type of
requests and size, their mask and anti-mask,
and linear or random addressing mode.
Then, we read the total number of accesses, maximum/minimum of read latency, and
aggregate read latency after 20 seconds for bandwidth and 200 seconds for thermal
measurements.
We calculate bandwidth by multiplying the number of accesses by the cumulative  
size of request and response packets \emph{including header, tail} and data payload (shown in 
Table~\ref{tab:hmc-rdwrsize}), and dividing it by the elapsed time.
Experiments are done with full-scale GUPS, unless stated otherwise.

\par
\textbf{\emph{Small-scale GUPS:}}
To measure latency-bandwidth relationships, we use a variation
of full-scale GUPS, small-scale GUPS, in which we change the number of active
ports to tune the request bandwidth. Similar to full-scale GUPS,
we perform each experiment for 20 seconds.

\par
\textbf{\emph{Stream GUPS:}}
We use Xilinx's AXI-Stream interface to send/\-receive
a group of requests/responses through streams to each port.
Stream interface provides fast and efficient communication 
between the host and the FPGA.
With stream GUPS, we also confirm the data integrity of our writes and reads.
To measure low-load latency, we use stream GUPS, so
that we can tune the number of requests.

\section{Experimental Results}
\label{results}

%
%-------------------------------------------------------------------------%
\subsection{Address Mapping Experiments}
\label{sec:res-address}

\begin{figure}[t]
\centering
\vspace{-0.1in}
\includegraphics[width=1\linewidth]{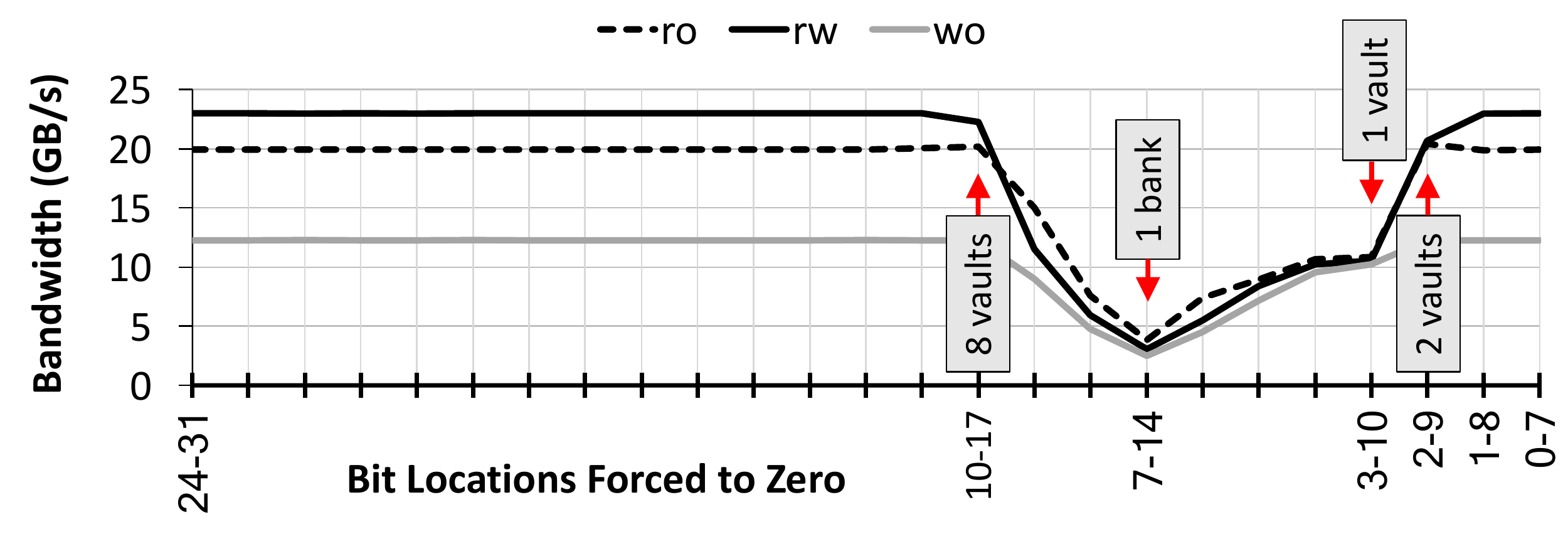}
\vspace{-0.1in}
\captionsetup{singlelinecheck=on,aboveskip=-2pt,belowskip=-3pt}
\caption{Applying an eight-bit mask to various bit positions of addresses. 
Similar to all following experiments, two half-width links are active, and raw
bandwidth, including header and tail, is reported.
}
\label{fig:address-map-exp}
\vspace{-0.19in}
\end{figure}

\noindent
To characterize the bandwidth properties of HMC, 
we explore a range of address space
mappings to its abstraction hierarchy (i.e., quadrants, vaults,
and banks) by generating random accesses
throughout the address space and evaluating the achieved bandwidth.
Address space mappings are achieved by starting with the
default low-order interleaved address
space mapping with 128\,B blocks (as described in Section~\ref{sec:hmc-add})
and then applying an eight-bit mask, which forces eight bits of address to \texttt{0}.
By applying the mask to various bit positions we map
the address space across quadrants, vaults and banks.
Figure~\ref{fig:address-map-exp} shows measured bandwidth
for three request types of \texttt{ro},
\texttt{wo}, and \texttt{rw}
with the size of 128\,B.
In this figure, bandwidth is the lowest when the mask is applied to bits 7-14,
which forces all references to bank 0 of vault 0 in quadrant 0 (Figure~\ref{fig:hmc-address}a).
The bandwidth rises with masking higher bits after the lowest
bandwidth point because
we incrementally spread the requests over more vaults, 
quadrants, and banks
with the masks after that.
Bandwidth experiences a large drop from masking bits 2-9 to
the masking bits 3-10 in \texttt{ro} and \texttt{rw} 
because for the latter mask the accesses
are restricted to a single vault with maximum internal
bandwidth of 10\,GB/s~\cite{ros14}.
In the rest of the paper, based on the results of this section,
we create targeted access patterns.
For instance, a 2-bank access pattern targets two banks 
within a vault, and a 4-vault access
pattern targets all the banks within four vaults.
%
%
%
%
%
%
%
%-------------------------------------------------------------------------%
%-------------------------------------------------------------------------%
\subsection{Bandwidth Experiments}
\label{sec:res-bw}

\begin{figure}[b]
\centering
\vspace{-0.18in}
\includegraphics[width=1\linewidth]{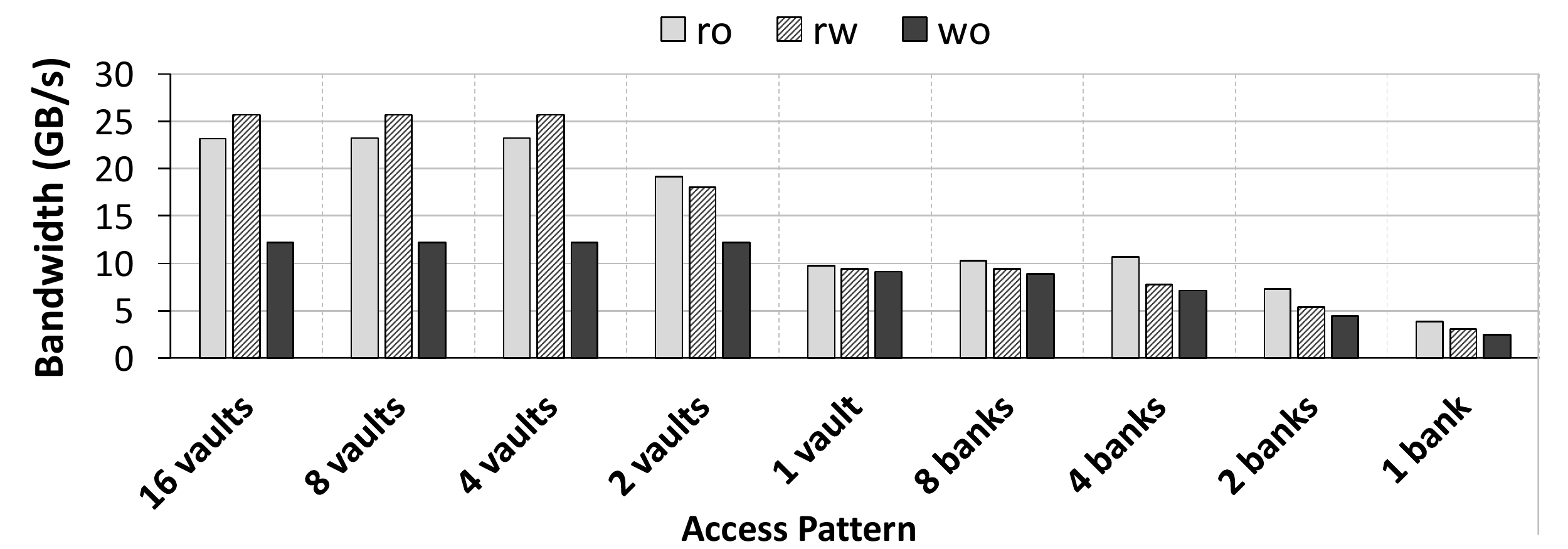}
\vspace{-0.25in}
\captionsetup{singlelinecheck=on,aboveskip=10pt,belowskip=0pt}
\caption{Measured HMC bandwidth for different types of accesses:
\texttt{ro}, \texttt{wo}, \texttt{rw}. The data size of each access
is 128\,B, or 8\,flits.}
\label{fig:res-bw}
\vspace{-0.05in}
\end{figure}

Figure~\ref{fig:res-bw} illustrates the bandwidth of various access
patterns.
As we pointed out in Section~\ref{sec:res-address}, since the bandwidth of a vault is limited
to 10\,GB/s, we observe that accessing more than eight banks of a vault does not affect the
bandwidth for all types of accesses.
For distributed accesses, in which addresses are spread across the
address space (e.g., across vaults and banks),
bandwidth of \texttt{rw} is higher than that of \texttt{ro}
because as read-modify-write operations consists of a read 
followed by a write,
both its request and response include data payloads, 
so \texttt{rw} utilizes bi-directional
links more effectively.
Meanwhile, as reads and writes are not independent,
the number of reads are limited by the number of writes.
Therefore, the bandwidth of \texttt{rw} is roughly double the bandwidth of \texttt{wo}.
\begin{figure}[t]
\centering
\vspace{-0.1in}
\includegraphics[width=1\linewidth]{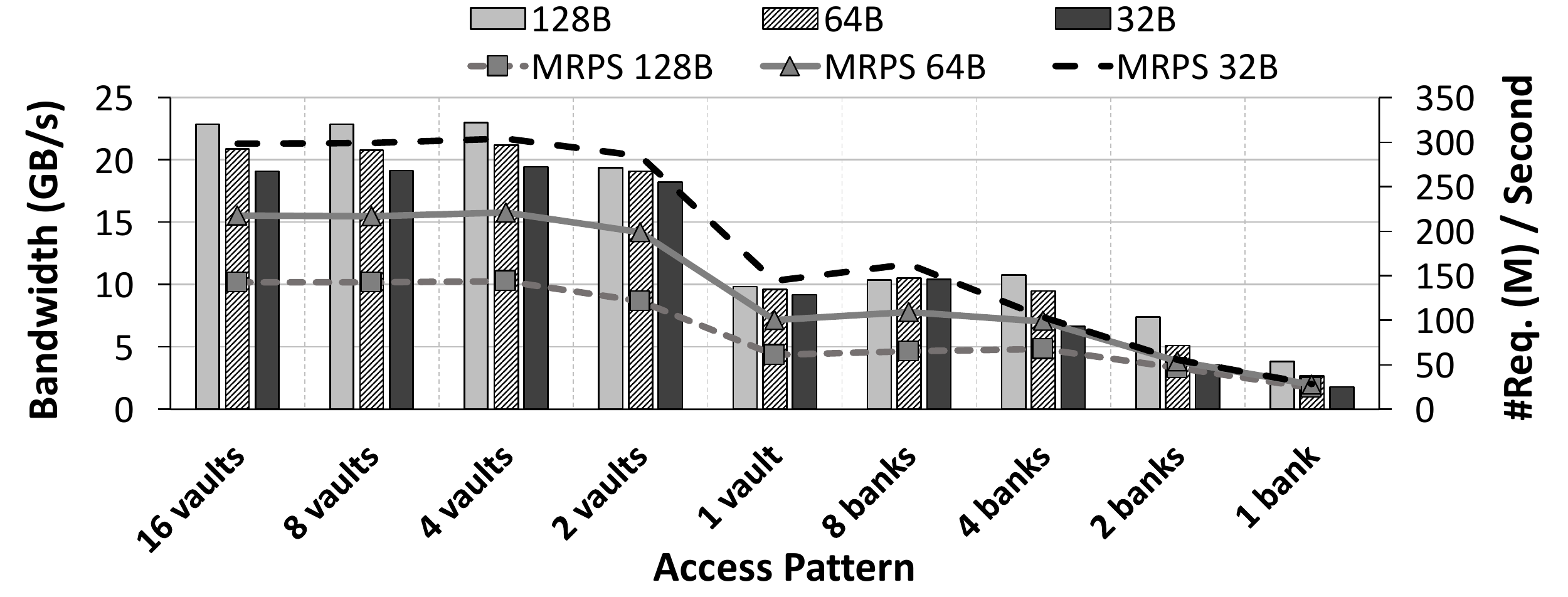}
\vspace{-0.25in}
\captionsetup{singlelinecheck=on,aboveskip=10pt,belowskip=0pt}
\caption{Measured HMC bandwidth for read-only requests with different request
sizes.
Lines represent the number of sent requests in millions per second (MRPS).}
\label{fig:res-bw-size}
\vspace{-0.24in}
\end{figure}

Figure~\ref{fig:res-bw-size} depicts the bandwidth of \texttt{ro} with 
the request sizes
of 128, 64, and 32\,B. As discussed in Section~\ref{sec:hmc-add},
the granularity of the DRAM data bus within a vault is 32\,B, and DRAM 
row size is 256\,B.
Therefore, reading the values smaller than 32\,B requires one data transfer, 
and reading
values larger than 32\,B requires multiple data transfers.
Note that since the maximum request size is smaller than the DRAM row size and
HMC follows the closed-page policy, each request incurs the
cost of opening just one DRAM row.
(We investigate the effects of closed-page policy in Section~\ref{sec:res-close}.)
Figure~\ref{fig:res-bw-size} also presents million requests
per second (MRPS) for each data size.
Although for less distributed access patterns (e.g., two banks) we see a
similar number of requests,
for distributed access patterns (e.g., 4, 8, and 16 vaults),
the MRPS differs in various request sizes.
For instance, when accesses distributed over 16 vaults, for small request sizes such as 32\,B,
HMC handles twice as many
requests as large request sizes such as 128\,B.
This behavior in the number of requests handled by HMC, combined with
the relatively same bandwidth utilization for various request sizes,
demonstrates that the bottleneck in bandwidth
is limited by DRAM timings and communication bandwidth,
rather than the size of components in the FPGA such
as GUPS buffers, tag-pool,
or in internal memory controllers within vaults.
In addition to request sizes shown in Figure~\ref{fig:res-bw-size}, we
perform the same experiments with all possible request 
sizes\footnotemark\,and discover that the trend in both bandwidth and
MRPS is similar.
\footnotetext{Possible request sizes are: 16, 32, 48, 64, 80, 96, 112, and 128\,B.
Note that for each request, we have an overhead of 16\,B as well. Therefore, as size increases,
the efficiency of data size to overhead also increases.}
%
%
%
%
%
%

%-------------------------------------------------------------------------%
%-------------------------------------------------------------------------%
\subsection{Thermal and Power Experiments}
\label{sec:res-th-pow}

\noindent

\begin{figure}[t]
    \centering
    \vspace{-0.1in}
    \begin{subfigure}[b]{0.5\textwidth}
        \includegraphics[width=1.0\textwidth]{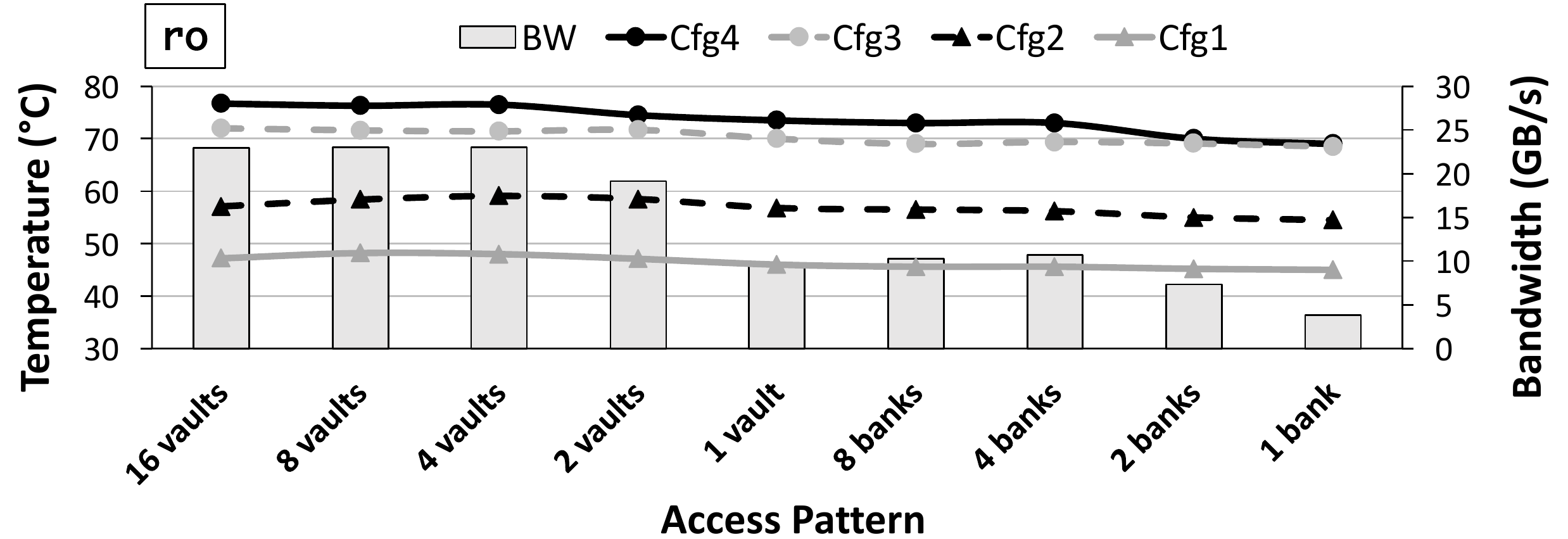}
        \centering
        \captionsetup{singlelinecheck=on,aboveskip=0pt,belowskip=0pt}
        \caption{}
        \label{fig:th-ro}
    \end{subfigure}
    \begin{subfigure}[b]{0.5\textwidth}
        \includegraphics[width=1.0\textwidth]{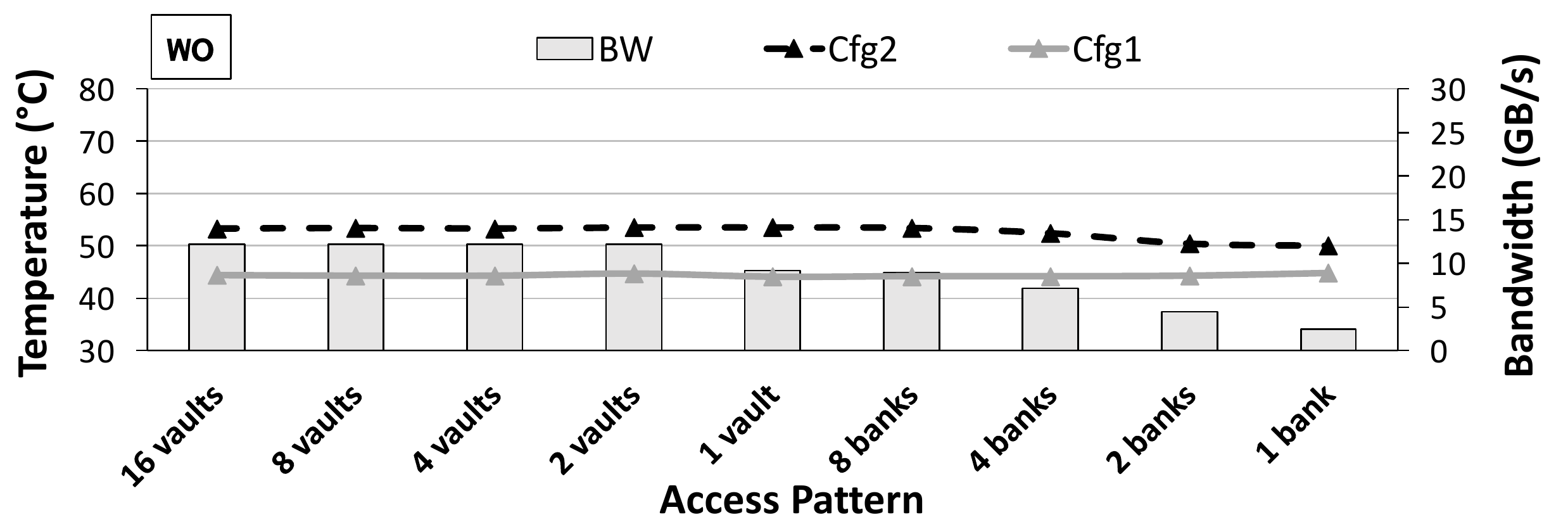}
        \centering
        \captionsetup{singlelinecheck=on,aboveskip=0pt,belowskip=0pt}
        \caption{}
        \label{fig:th-wo}
    \end{subfigure}

    %\rule{0.3\textwidth}{0.02em}
    \begin{subfigure}[b]{0.5\textwidth}
        \includegraphics[width=1.0\textwidth]{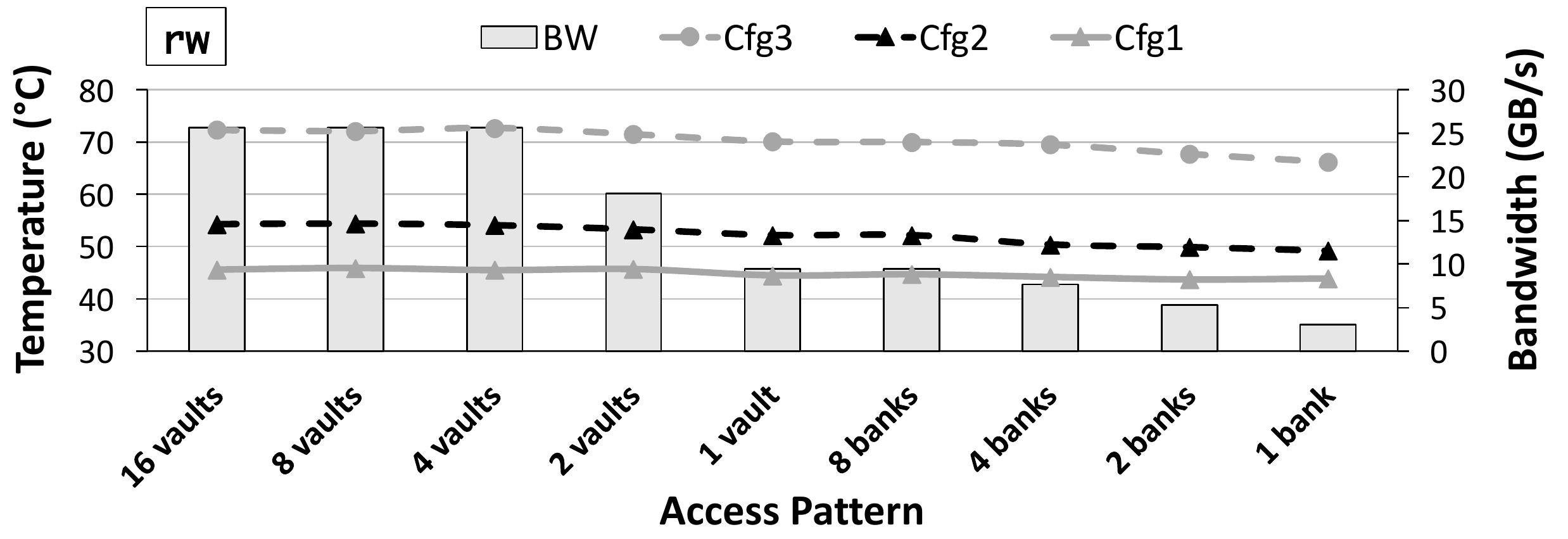}
        \centering
        \captionsetup{singlelinecheck=on,aboveskip=0pt,belowskip=0pt}
        \caption{}
        \label{fig:th-rw}
    \end{subfigure}

    \captionsetup{singlelinecheck=on,aboveskip=3pt}
    \caption{Heatsink surface temperature and bandwidth of HMC 
    during various access patterns and
    request types: (a) Read-only, (b) write-only, and (c) read-modify-write.}
    \label{fig:th-all}
    \vspace{-0.35in}
\end{figure}

In general, 3D systems present limiting thermal and power
challenges. 
Therefore, we are interested in understanding the
relationships between temperature, bandwidth, and power consumption
particularly considering that the high speed links are responsible for
significant power consumption. 
We are specially interested in the
perspective of future PIM architectures that will
create higher temperatures. 
Therefore, we push operation to the
thermal limits by creating high-temperature environments through
manipulation of the cooling configurations in Table~\ref{tab:cooling}
to study the relationship between temperature and bandwidth. 
We use
synthetic address generation to explore these limits, and  follow the methodology presented in Section~\ref{sec:meth-firm}
for temperature and power experiments.  
Figure~\ref{fig:th-all} shows the outcomes of thermal
experiments for cooling configurations shown in
Table~\ref{tab:cooling}.  
(This figure does not include configurations
that trigger HMC failures.) 
Although experiments
are limited by the resolution of the experimental 
infrastructure ($\pm 0.1$\textdegree C),
the absolute temperature values provide a close approximation for
temperature-related behaviors.
The first observation is that the behavior of HMC temperature (i.e.,
the reduction of temperature when bandwidth decreases) is similar in
all cooling environments.  
That is, during the first three access
patterns (accessing 16 to four vaults), 
in which bandwidth utilizations
are similar, temperature remains constant.  
Then, during the
execution of the rest of the access patterns (accessing two vaults to
one bank), in which the bandwidth utilization decreases, temperature
drops.  
In fact, more bandwidth utilization causes more accesses
in DRAM layers, processes in memory controller, 
and data transfers in SerDes circuits, which consume
43\% of total power~\cite{pug:jes14,
jed:kee12, nai:had17}.  
As bandwidth utilization increases, HMC
temperature rises, which directly affects the power consumption.
Note that the bandwidth profile is \emph{directly} affected by access patterns.
Since HMC, a 3D-stacked design, has a smaller size than
traditional DRAM DIMMs and has the capability to achieve high
bandwidth (by exploiting serial communication and high BLP), 
we conclude that HMC and generally 3D-stacked memories
are susceptible to thermal bottleneck.
As mentioned, previous work assumed that the reliable temperature
bound of DRAM is $85$\textdegree C~\cite{zhu:yux:2016,yas:nuw14}.
To distinguish the operational temperature for HMC,
we run several experiments, forcing higher thermal fields by
controlling  the cooling configurations
during which failures occur around the same temperatures.
Operationally, to indicate an inevitable thermal failure (i.e.,
shutdown), HMC uses the head/tail of response messages to transmit
this information to the host.  
Occurrence of a thermal failure during
an experiment stops the HMC and recovery from it entails following
steps: Cooling down, resetting HMC, resetting FPGA modules such as
transceivers, and initializing HMC and FPGA.  
Notice that when failure
occurs, stored data in DRAM is lost. Data recovery must rely on
external techniques such as checkpointing and rollback.
During our experiments, we observe that read-only accesses do not
experience any thermal failures, even with the lowest cooling
environment in \texttt{Cfg1}, in which temperature reaches to
$80$\textdegree C.  
However, for \texttt{wo} and \texttt{rw}
requests, the same cooling environment (\texttt{Cfg1}) triggers
thermal failures.  
Hence, our experiments indicate that reliable
temperature bound is lower for workloads with significant write
content - around $75$\textdegree C, which is about $10$\textdegree C
lower than that for read-intensive accesses.
%
% >>>>>>> Conclusion of Temperature/BW by referring to figure 10 A
%

\begin{figure}[t]
\centering
\vspace{-0.10in}
\includegraphics[width=1\linewidth]{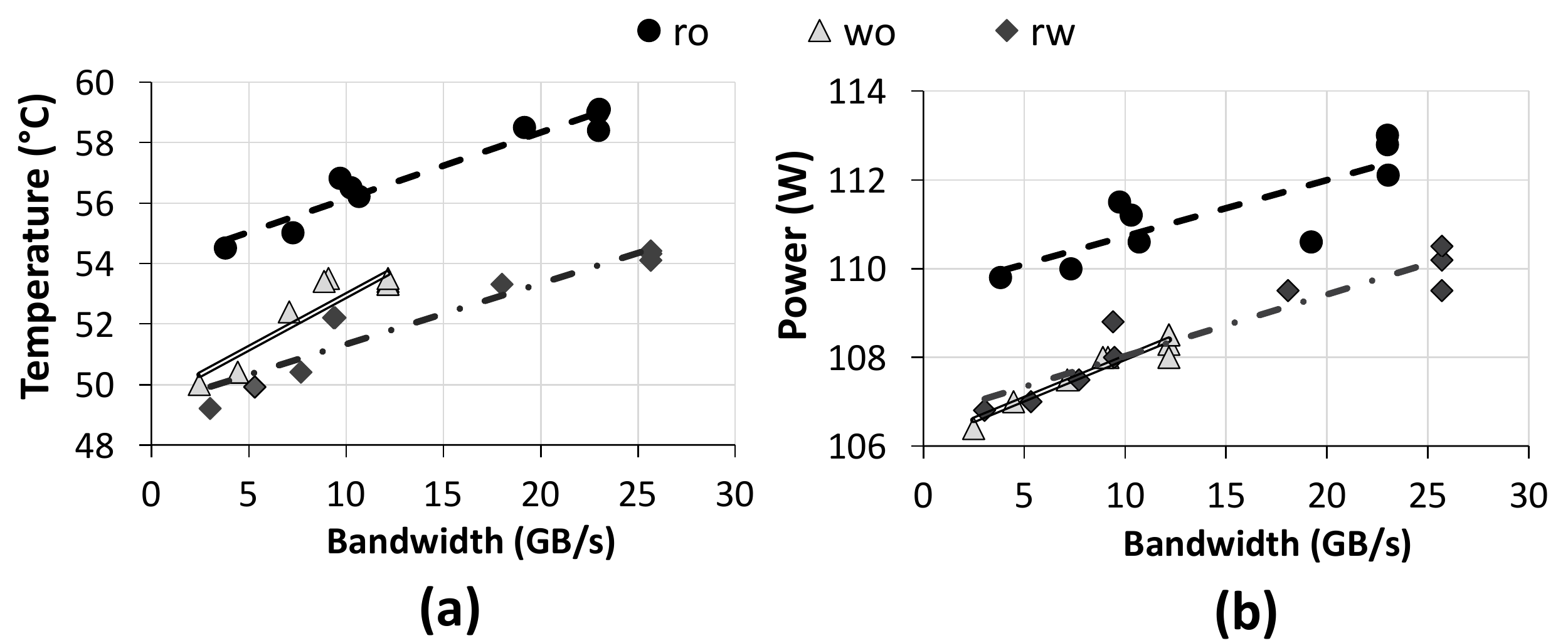}
\vspace{-0.22in}
\caption{ (a) Temperature and (b) power consumption relationships with bandwidth in \texttt{Cfg2} 
for different request types.}
\label{fig:bw-line-2x}
\vspace{-0.2in}
\end{figure}

The relationship between the bandwidth and temperature,
shown conceptually in Figure~\ref{fig:moti}a, is drawn in
Figure~\ref{fig:bw-line-2x}a with a linear regression fit to measured
data.  
Figure~\ref{fig:bw-line-2x} is extracted based
on results in \texttt{Cfg2}, a configuration with highest
temperature, in which none of the \texttt{ro}, \texttt{wo}, or
\texttt{rw} experienced any thermally induced failures, providing a
fair comparison.  
As this graph illustrates, when bandwidth increases from five to 20\,GB/s, the temperature rises
$3$\textdegree C and $4$\textdegree C for \texttt{ro} and
\texttt{rw} accesses respectively.
The positive slope of all three
lines in the graph illustrates that the temperature bottleneck with
increasing bandwidth is inevitable in such 3D memory organizations and therefore this 
characterization is important.  
Also, the greater slope of the line
corresponding to \texttt{wo} indicates that write operations are more
temperature sensitive across this bandwidth range.  
We could not assert the reason behind this with our current
infrastructure.  
However, the results suggest that such
sensitivity favors read-intensive workloads such as streaming
applications to maximize bandwidth capacity.   

\begin{figure}[t]
    \centering
    \vspace{-0.1in}
    \begin{subfigure}[b]{0.5\textwidth}
        \includegraphics[width=1.0\textwidth]{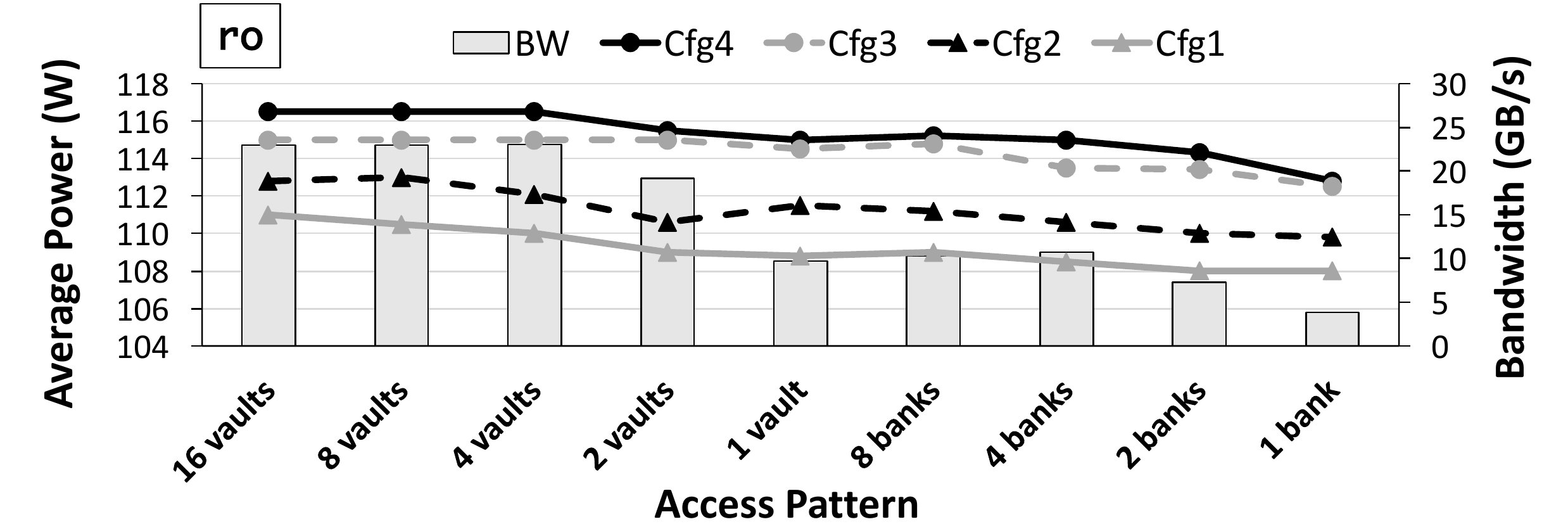}
        \centering
        \captionsetup{singlelinecheck=on,aboveskip=0pt,belowskip=0pt}
        \caption{}
        \label{fig:pow-ro}
    \end{subfigure}

    %\rule{0.3\textwidth}{0.02em}
    \begin{subfigure}[b]{0.5\textwidth}
        \includegraphics[width=1.0\textwidth]{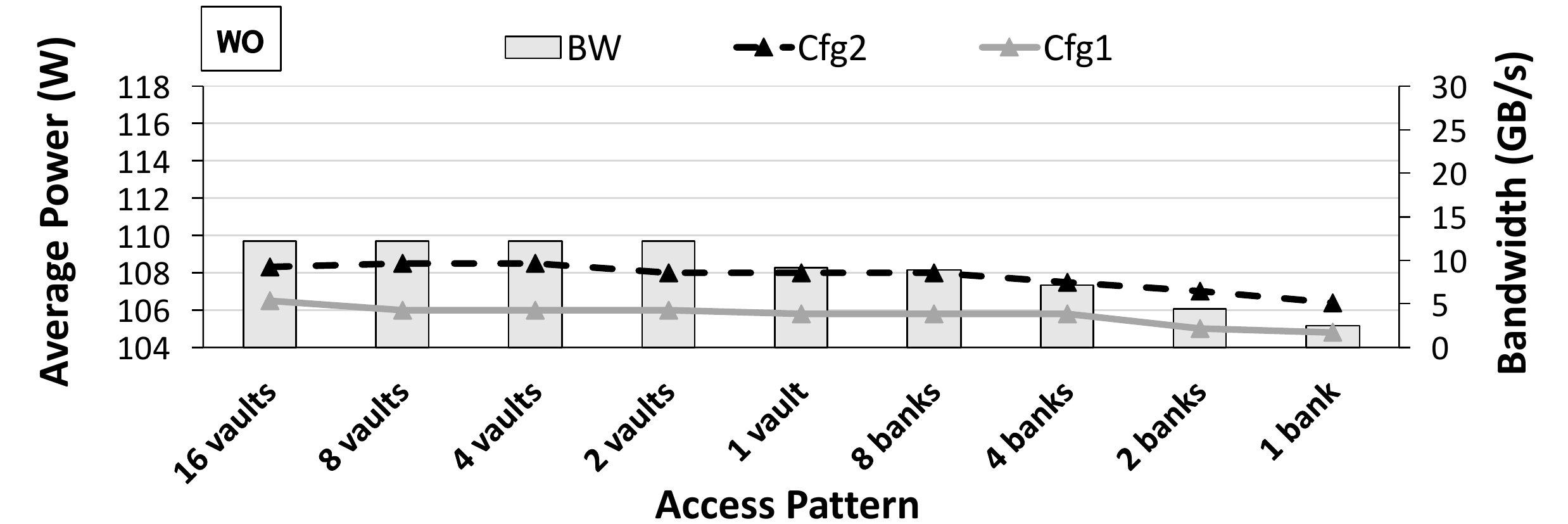}
        \centering
        \captionsetup{singlelinecheck=on,aboveskip=0pt,belowskip=0pt}
        \caption{}
        \label{fig:pow-wo}
    \end{subfigure}

    %\rule{0.3\textwidth}{0.02em}
    \begin{subfigure}[b]{0.5\textwidth}
        \includegraphics[width=1.0\textwidth]{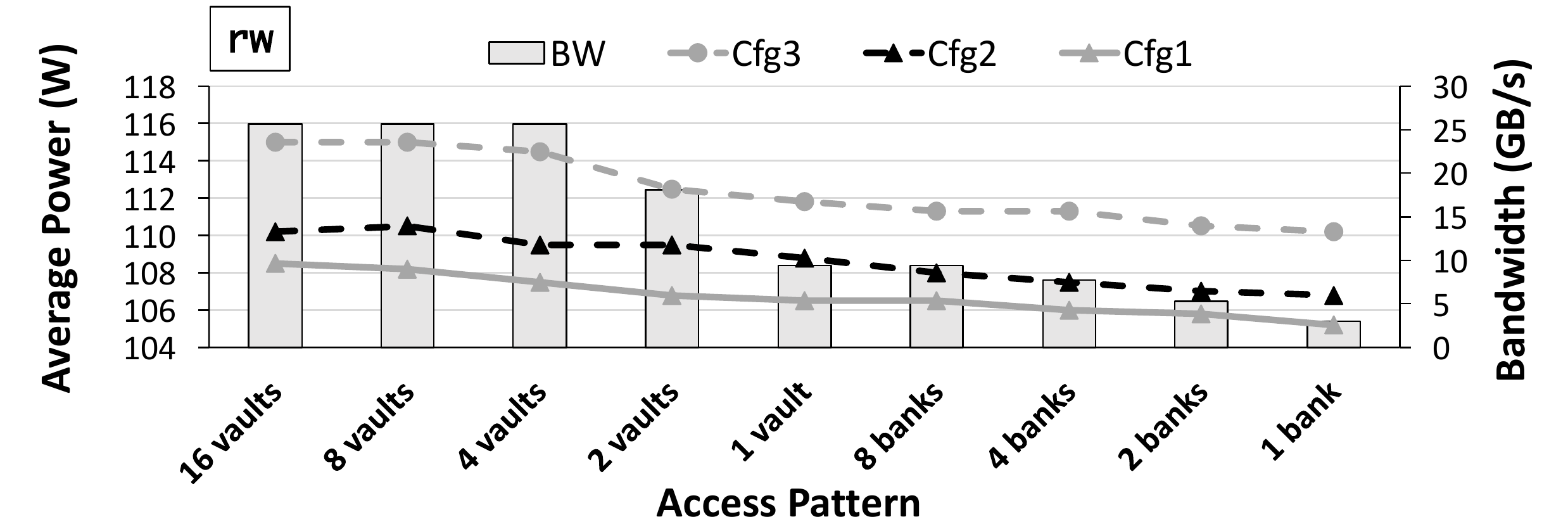}
        \centering
        \captionsetup{singlelinecheck=on,aboveskip=0pt,belowskip=0pt}
        \caption{}
        \label{fig:pow-rw}
    \end{subfigure}
    
    \captionsetup{singlelinecheck=on,aboveskip=4pt}
    \caption{Average power consumption of the system and HMC bandwidth
    during various access patterns and
    request types: (a) Read-only, (b) write-only, and (c) read-modify-write.}
    \label{fig:pow-all}
    \vspace{-0.35in}
\end{figure}
\begin{figure}[t]
\centering
\vspace{0.05in}
\includegraphics[width=1\linewidth]{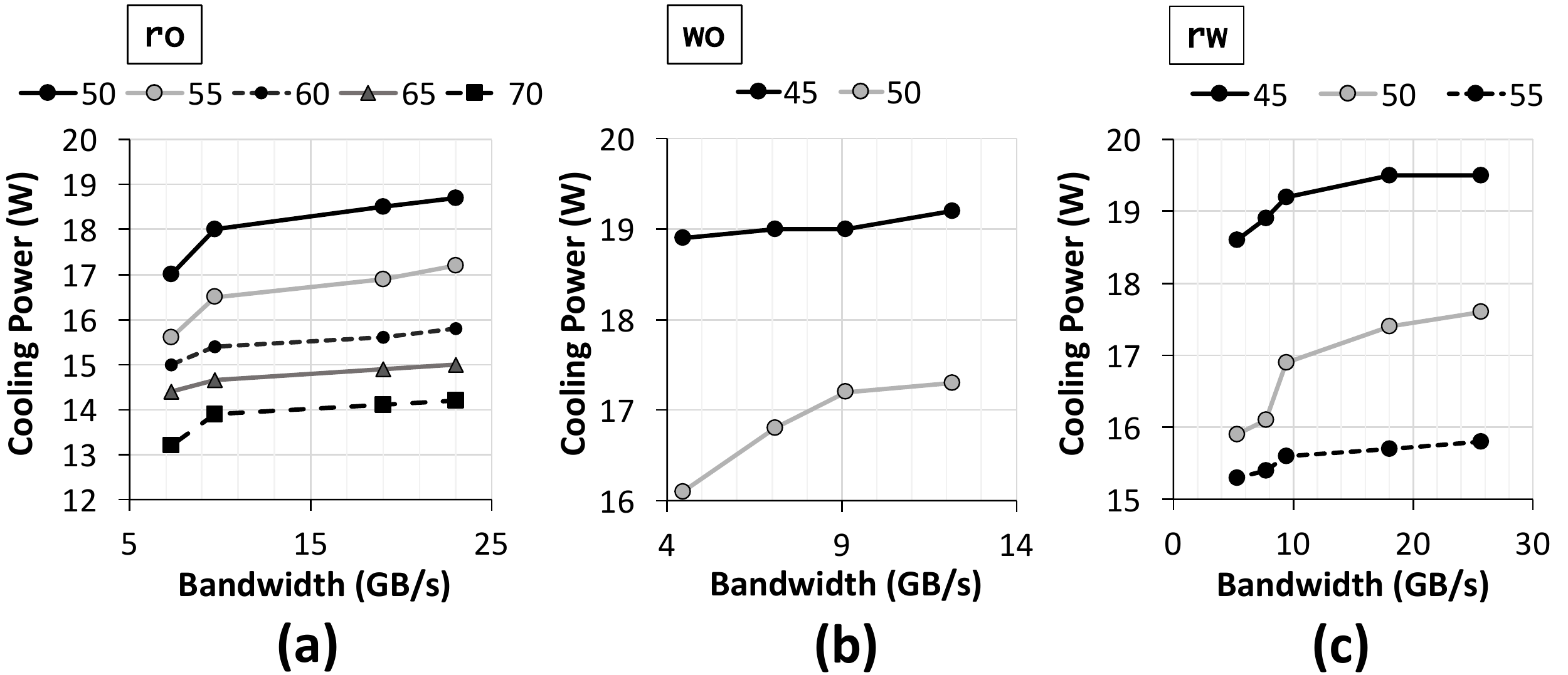}
\vspace{-0.22in}
\caption{Observed cooling power and bandwidth relationships for three access types.
Each line indicates cooling power required for maintaining the system at a specific
temperature.}
\label{fig:cooling-power}
\vspace{-0.3in}
\end{figure}

Figure~\ref{fig:pow-all} illustrates power consumption behavior across
request types and distributions. 
First, as one might expect power
consumption increases with increased bandwidth. 
Second, decreased cooling capacity leads to higher power consumption for the same
bandwidth reflecting the coupling between power and temperature. This
is particularly important to be aware of for PIM
configurations.
Figure~\ref{fig:bw-line-2x}b provides a linear regression fit
line based on data from  Figure~\ref{fig:pow-all}.
The figure illustrates a 2\,W increment in device power consumption
when bandwidth utilization rises from five to 20\,GB/s.
%
%------------Cooling Power------------
%
%
However, cooling capacity is also another contributing factor to power consumption.
In fact, the cooling solution that is required to maintain a device at
a reliable temperature, can consume significant power, and we study
the impact as follows. 
With respect to voltage and current, listed in
Table~\ref{tab:cooling} for two PCIe backplane fans, and by the fact that
as distance increases, the effective power consumption for cooling of
the external fan decreases, we calculated the power consumption of each cooling
configurations as 19.32, 15.9, 13.9, and 10.78\,W
for \texttt{Cfg1} to \texttt{Cfg4}, respectively.  
Then, by using linear regression over measured data of temperature
and bandwidth (Figure~\ref{fig:th-all}), we extract the graphs in
Figure~\ref{fig:cooling-power}, which reflect the conceptual
behaviors shown in Figure~\ref{fig:moti}b.
In Figure~\ref{fig:cooling-power}, each line shows required cooling power for
maintaining a specific temperature as bandwidth utilization increases.
Note that the absolute values of cooling power are related to our cooling
infrastructure.
However, Figure~\ref{fig:bw-line-2x}b and \ref{fig:cooling-power}
indicate that as bandwidth utilization
increases, power consumption of both the device and the cooling
solution increases. 
In average, an increase of 16\,GB/s in
bandwidth causes a growth of 1.5\,W in cooling power.
Also, as the peak temperature increases exponentially with the proximity 
of the compute unit~\cite{zhu:yux:2016}, adding PIM-like capabilities in the logic layer significantly increases the cost of cooling.

%
%
%
%

%-------------------------------------------------------------------------%
%-------------------------------------------------------------------------%
\subsection{Closed-Page Policy Experiments}
\label{sec:res-close}

\begin{figure}[b]
\centering
\vspace{-0.1in}
\includegraphics[width=1.0\linewidth]{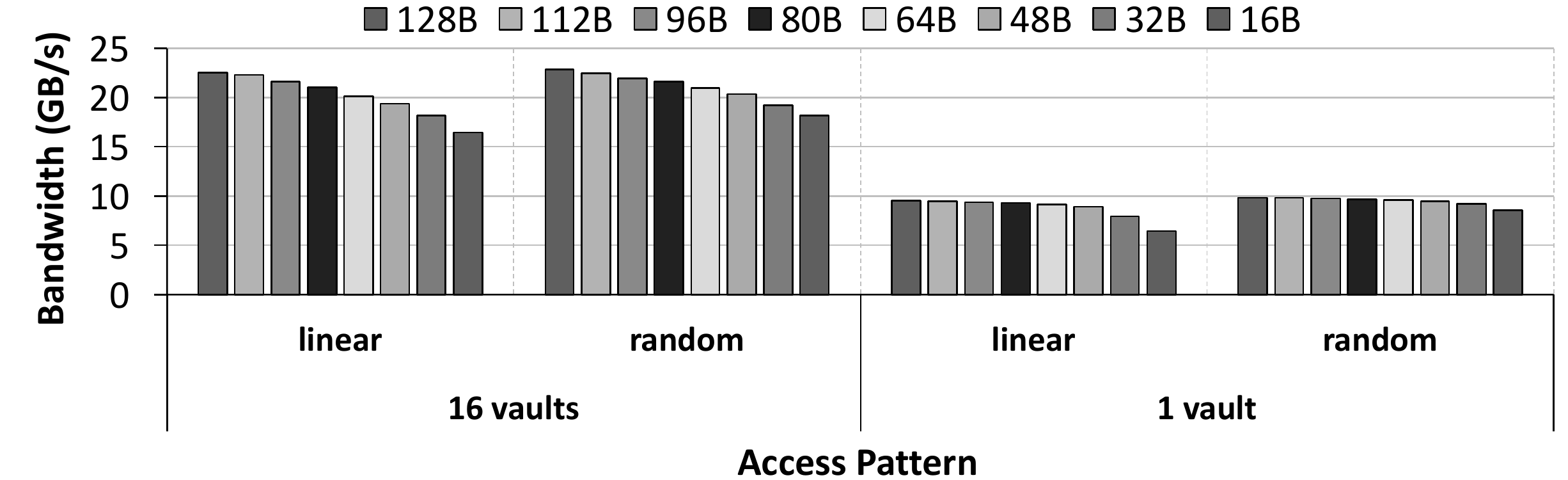}
\centering
\vspace{-0.2in}
\caption{Measured HMC bandwidth for random and linear
read-only requests with different request sizes.}
\label{fig:res-bw-open}
\vspace{-0.15in}
\end{figure}
\noindent
As discussed in Section~\ref{sec:hmc-add}, the HMC implements a
closed-page policy.  While with the open-page policy, the average
latency of linear accesses is lower than that of random accesses because
of row buffer hits; with a closed-page policy, the average
latencies of linear and random accesses should be the same.  To
illustrate this, we perform an experiment by issuing linear and random
addresses and measure the bandwidth utilization.  As
Figure~\ref{fig:res-bw-open} illustrates, the bandwidth of random and
linear accesses are similar (random accesses, compared to linear accesses, 
have slightly higher bandwidth because of fewer conflicts on any
of the shared resources). 
We conjecture that the increase in
bandwidth from 32\,B blocks to 128\,B blocks is due to efficiencies in
streaming a block size greater than the width of the 32\,B DRAM bus and
reduced packet overhead due to the larger data size.
We observe that applications benefit from increasing parallelism
over vaults and banks (e.g., stripe data across the vaults and banks), 
rather than increasing reference locality.  For
instance, a streaming application that exhibits linear references
should not allocate data sequentially within a vault for two reasons:
(\romannum{1}) The maximum internal bandwidth of a vault is 10\,GB/s, which
limits the total bandwidth; and (\romannum{2}) references to successive addresses
cannot exploit spatial locality because of the closed-page policy.
Furthermore, to increase the
effective bandwidth, the application should use 128\,B size for
requests because each request/response has an overhead of one
flit, so a larger data size utilizes the available bandwidth more
effectively.
For instance, when using 128\,B request size, the effective bandwidth is 
$\nicefrac{128\text{B}}{(128B+16\text{B})}=89\%$ of raw bandwidth.
However, when using 16\,B request size, the effective bandwidth is 
$\nicefrac{16\text{B}}{(16\text{B}+16\text{B})}=50\%$ of raw bandwidth.
Therefore, the OS, memory controller, or programmer has to promote
parallelism, remapping of data, and concatenation of requests to 
achieve high effective bandwidth.

%
%
%
%

%-------------------------------------------------------------------------%
\subsection{Latency Experiments}
\label{sec:res-lat}
\noindent
As HMC, a 3D-stacked design, exploits packet-switched interfaces for scalability and PIM features, it yields another important performance attribute, latency, a less-focused attribute in traditional DIMM memories because of their constant timing characteristics.
Thus, we dedicate this section to latency experiments and deconstruction of its contributing factors.

%-------------------------------------------------------------------------%
\emph{\textbf{1) Contributing factors:}}
Each port in the GUPS (Section\-~\ref{sec:meth-firm}) measures
the read latency as the number of cycles from when a read request is
submitted to the HMC controller (on the FPGA) until the port receives
the read response.
This latency includes cycles for
(1) arbitrating among ports,
(2) creating a packet
(i.e., generating head and tail flits, converting to flits,
adding sequence numbers, and generating CRCs),
(3) performing flow control,
(4) converting the packet to the SerDes protocol~\cite{hybrid2013hybrid1},
(5) serializing the packet,
(6) transmitting the packet on the links,
(7) processing the packet in the HMC (or memory) and generating a response,
(8) transmitting the response back,
(9) deserializing the response,
(10) performing necessary verifications
(i.e., checking errors and sequence numbers), and
(11) routing the response back.
Although conventional memory access latency for DDR memories is
only reported by including items
6 to 8, a comprehensive latency should also include
the number of cycles of similar related items.\footnotemark
\footnotetext{The reason for that items from 6 to 8 are
reported in the JEDEC is that the standard specifies deterministic latencies,
and the memory controller is on the host side. Therefore,
memory vendors are not responsible for extra latencies.}
In addition, since packet communication enables features such as data
integrity, remaining items might have noticeable latencies, so
reporting only the items from 6 to 8 does not faithfully represent
latency.

\begin{figure}[t]
\centering
\vspace{-0.1in}
\includegraphics[width=0.9\linewidth]{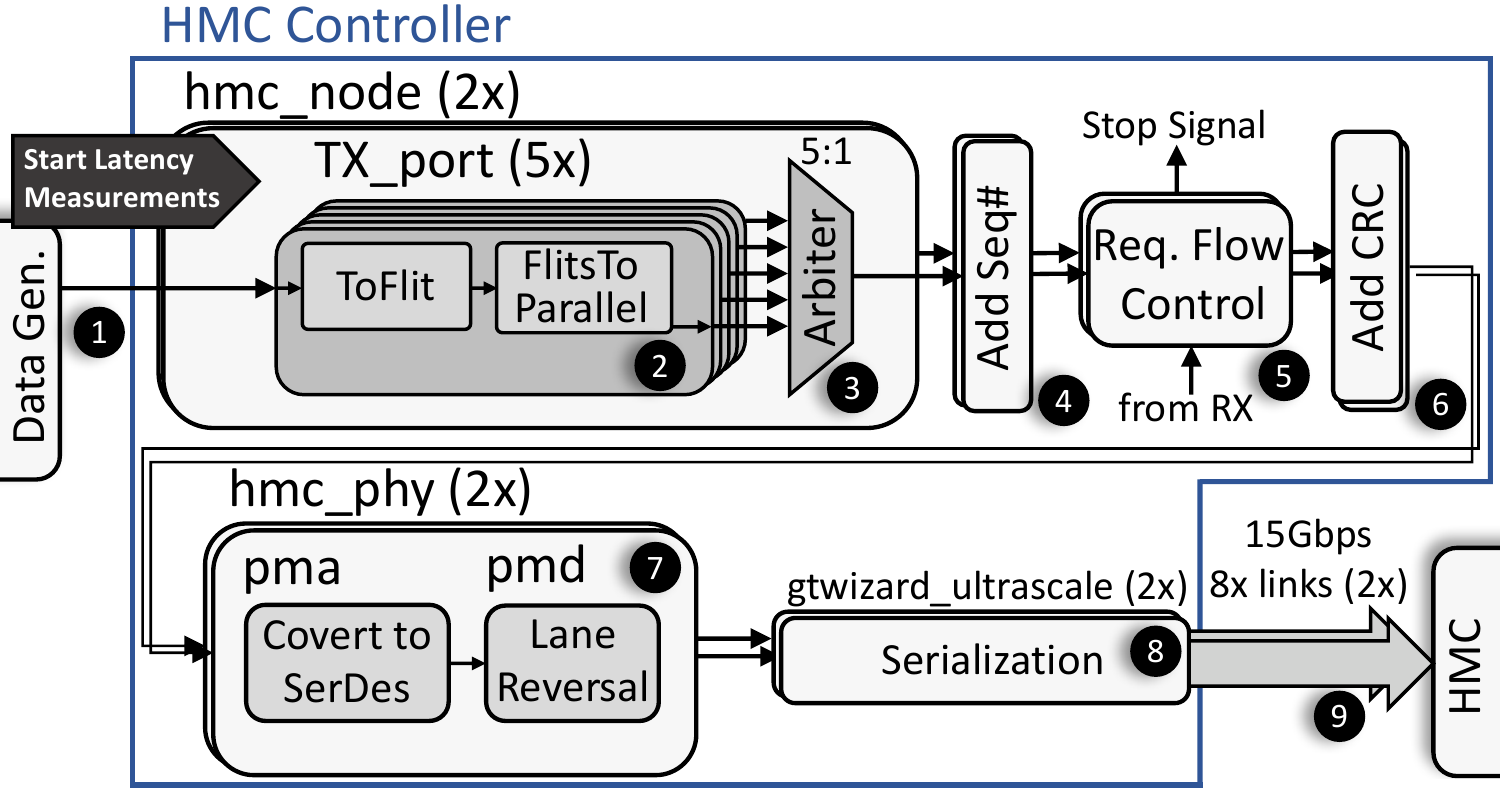}
\vspace{-0.1in}
\captionsetup{singlelinecheck=on,aboveskip=9pt,belowskip=4pt}
\caption{Deconstruction of transmit path (TX) in
the HMC controller module of Figure~\ref{fig:infs}b.}
\label{fig:decons}
\vspace{-0.35in}
\end{figure}
To understand the factors which contribute to latency,
we inspect the latencies associated with each module in the transmit (TX)
and receive (RX) paths by time stamping requests and reading the stamps
from each module.
Figure~\ref{fig:decons} presents the latency deconstruction of the TX path after a request is submitted to the HMC controller. Each external link is connected to an \texttt{hmc\_node} module, which consists of five \texttt{TX\_ports} that correspond to a port. Since the AC-510 has two links to the HMC, this means 10 ports are available on the FPGA, but one is reserved for system use. 
After a request is submitted to the HMC controller (\ding{182} in Figure~\ref{fig:decons}), the \texttt{TX\_port} unit converts it to flits and buffers up to five flits in the \texttt{FlitsToParallel} unit (\ding{183}), which takes ten cycles or 53.3\,ns (based on the max frequency of the FPGA at 187.5\,MHz). Then, flits from each port are routed to subsequent units in a round-robin fashion by an \texttt{arbiter} (\ding{184}). This routing latency is between two to nine cycles. Afterwards, \texttt{Add-Seq\#}, \texttt{Req. Flow Control}, and \texttt{Add-CRC} units contribute a latency of ten cycles (\ding{185}, \ding{186}, and \ding{187}, respectively). These units add fields in packets for reordering and data integrity. Moreover, if the number of outstanding requests exceeds a threshold, the request flow-control unit sends a stop signal to the corresponding port (\ding{186}) requesting a pause in the generation of memory access requests. For low-contention latency measurements, the request flow-control unit does not stop the transfer of any flit. Finally, flits are converted to the SerDes protocol~\cite{hybrid2013hybrid1} and serialized (\ding{188} and \ding{189}, respectively), which takes around ten cycles. In addition, transmitting one 128\,B request takes around 15 cycles (\ding{190}). Overall, up to 54 cycles, or 287\,ns, are spent on the TX path. Similarly, for a packet, 260\,ns are spent on the RX path. In total, 547\,ns of measured latency is related to link transfers and packet generation on the FPGA.
Note that this is the minimum latency and is incurred when the flow-control unit does not stall transfers. For high-load studies, queuing delays will substantially increase this latency measurement.

%--------------------------------------------------%

\begin{figure}[t]
\centering
\vspace{-0.1in}
\captionsetup{singlelinecheck=on,aboveskip=-9pt,belowskip=-2pt}
\includegraphics[width=1\linewidth]{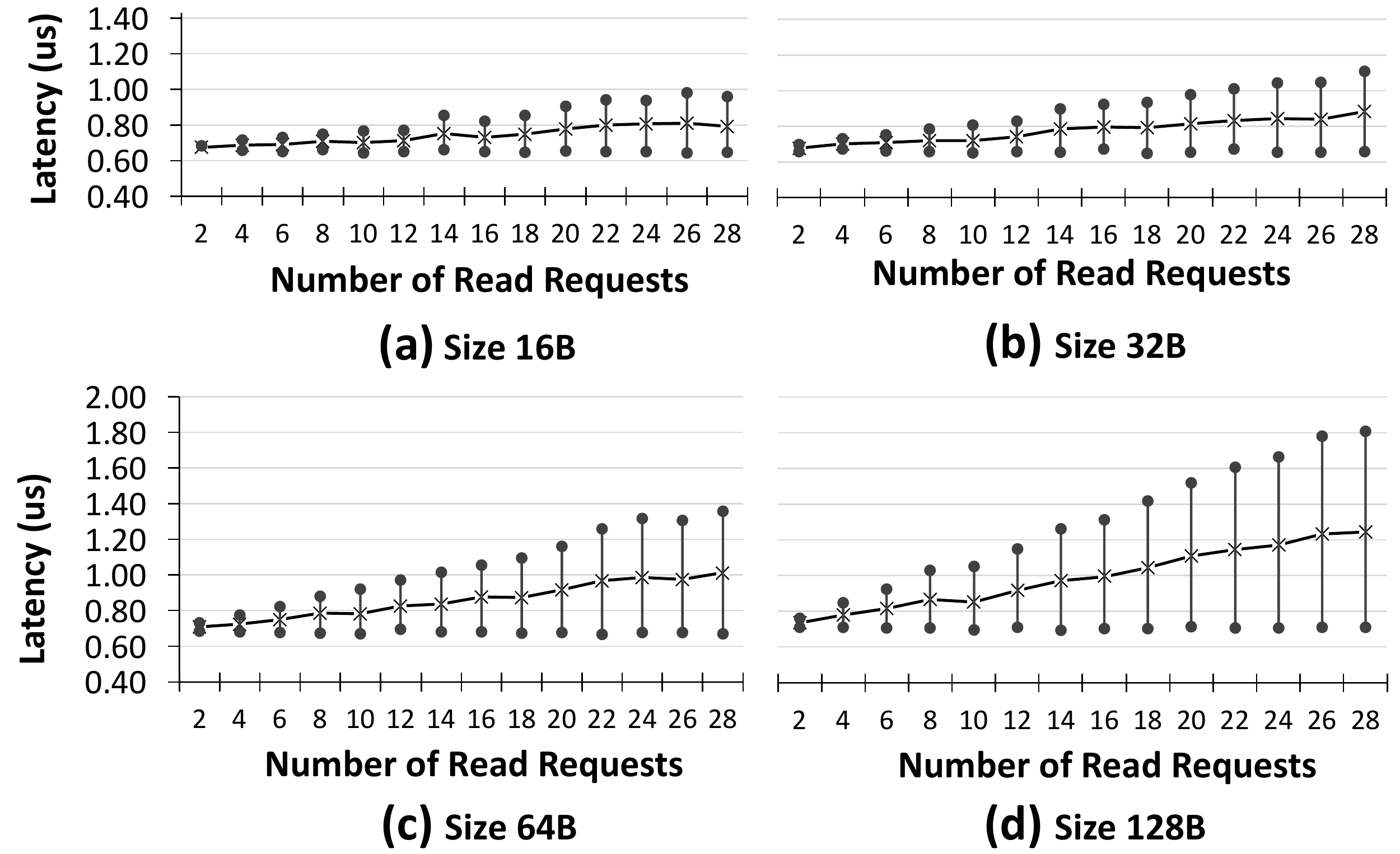}
\vspace{0in}
\caption{Average, minimum, and maximum latency of low-load accesses
for various request sizes.}
\label{fig:low-load-lat}
\vspace{-0.2in}
\end{figure}

\emph{\textbf{2) Low-load latency:}}
This paragraph explores the latency of low-load read accesses (i.e.,
no-load latency) and demonstrates the correlation between latency and
the number of read accesses for varied packet sizes.
As mentioned in Section~\ref{sec:meth-firm}, for these explorations, we use stream
GUPS.
Each subfigure of Figure~\ref{fig:low-load-lat} shows
variations in average, minimum, and maximum latency when the number of
read requests in a stream changes from two to 28.
This figure provides the following information about the behavior of low-load accesses.
First, a comparison between the slopes of four diagrams in this figure
reveals that when the request size is larger, latency increases faster
with regard to the number of read requests.
In other words, the latency of a stream of 28 large packets (i.e., 128\,B packets) is 1.5x
as high as that of a stream of 28 short packets (i.e., 16\,B packets).
However, a small stream (e.g., a stream of two packets) incurs almost
the same latency regardless of packet size.
Second, based on Figure~\ref{fig:low-load-lat}, 
as we expected, the increase in average
latency comes from increasing maximum latencies.
This is probably primarily
as a result of interference between 
packets in the logic layer which increases
with number of references and size of packets.
In addition, unlike maximum latencies, 
minimum latencies have a constant
value as the number of requests increases.
Third, the results for
various sizes show that the minimum latency of 128\,B packet sizes is
711\,ns, 56\,ns higher than that of 16\,B packet
sizes.
We should also add delay for transition across the
interface between a GUPS port and the HMC controller --- not expected to
be more than a few cycles in the low-load case.
These latencies contain
547\,ns infrastructure-related latency, so
in average 125\,ns is spent in the HMC.
The latency variation within HMC is due to 
packet-switched interface, and TSV and DRAM timings.
In comparison with typical DRAM access latency for 
a closed-page policy, we estimate the
latency impact of a packet-switched interface to be about two times
higher.
In return, we obtain higher bandwidths via high BLP, 
more scalability via
the interconnect, and better package-level 
fault tolerance via rerouting
around failed packages.

\emph{\textbf{3) High-load latency:}}
\begin{figure}[b]
\centering
\vspace{-0.15in}
\includegraphics[width=1\linewidth]{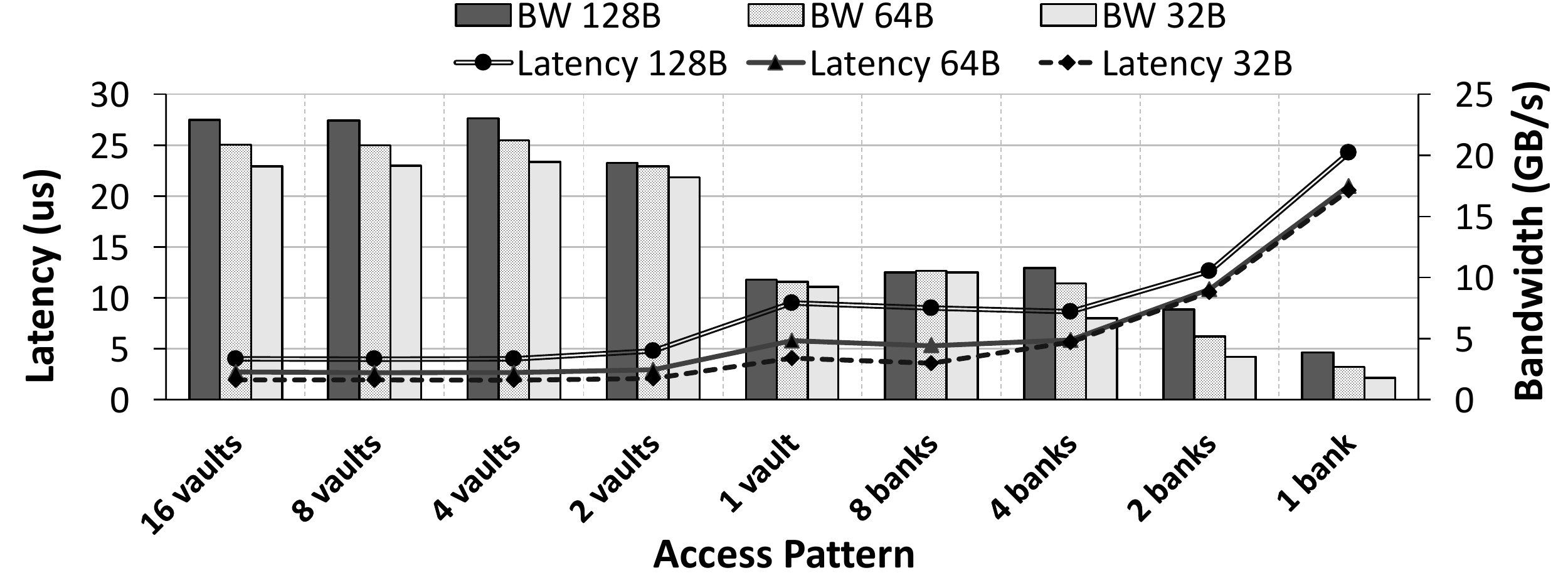}
\vspace{0in}
\captionsetup{singlelinecheck=on,aboveskip=-10pt,belowskip=0pt}
\caption{Measured read latency of high-load accesses for various access patterns and request sizes.}
\label{fig:res-latency}
\vspace{-0.05in}
\end{figure}
Figure~\ref{fig:res-latency} illustrates the latency behavior of
high-load accesses (\texttt{ro}).  
The read latency varies from
1,966\,ns for 32\,B requests spread across 16 vaults to 24,233\,ns for
128\,B requests targeted to a single bank.  
The average reference latency of a high-load access pattern is 12 times as high as that of a low-load access
pattern. 
These latency measurements are made at the data generation
component of a GUPS port and prior to the HMC controller. 
At high loads, reference requests are queued up at the HMC controller input and
we conjecture that the bulk of the growth in latency is due to this
queuing delay. 
This observation is supported by the fact that the
bandwidth utilization profile is similar across all experiments
described earlier.
Further, the latency of 32\,B read requests is always lower than that
of 64\,B and 128\,B read requests because the granularity of the DRAM data
bus within each vault is 32\,B, as pointed out in
Section~\ref{sec:hmc-add}.  
Therefore, the memory controller in the
logic layer has to wait a few more cycles when accessing data larger
than 32\,B.  
Figure~\ref{fig:res-latency} shows that latency is directly
related to bandwidth.  
A more distributed access pattern exploits both
multiple vault controllers and BLP.  
By contrast, a targeted access pattern to a vault or to the banks within a vault
incurs high latency because of the serialization of requests and
limited number of resources per bank.  
The micro-second
order of reported values for latency illustrates the importance of
understanding the internal structure and abstractions supported by
HMC, sensitivity to the internal frequency of 
operation,\footnotemark\, and
the importance of the design of the host-side memory interfaces.
\footnotetext{
  As Lee et al.~\cite{lee:ghose2016} pointed out, current 3D-stacked designs
  operate at lower frequencies than DDRs (e.g., Wide I/O~\cite{kim:chi2012}
  operates at 200-266\,MHz while DDR3 operates at 2,133\,MHz). }

\begin{figure}[t]
\centering
\vspace{-0.1in}
\includegraphics[width=1\linewidth]{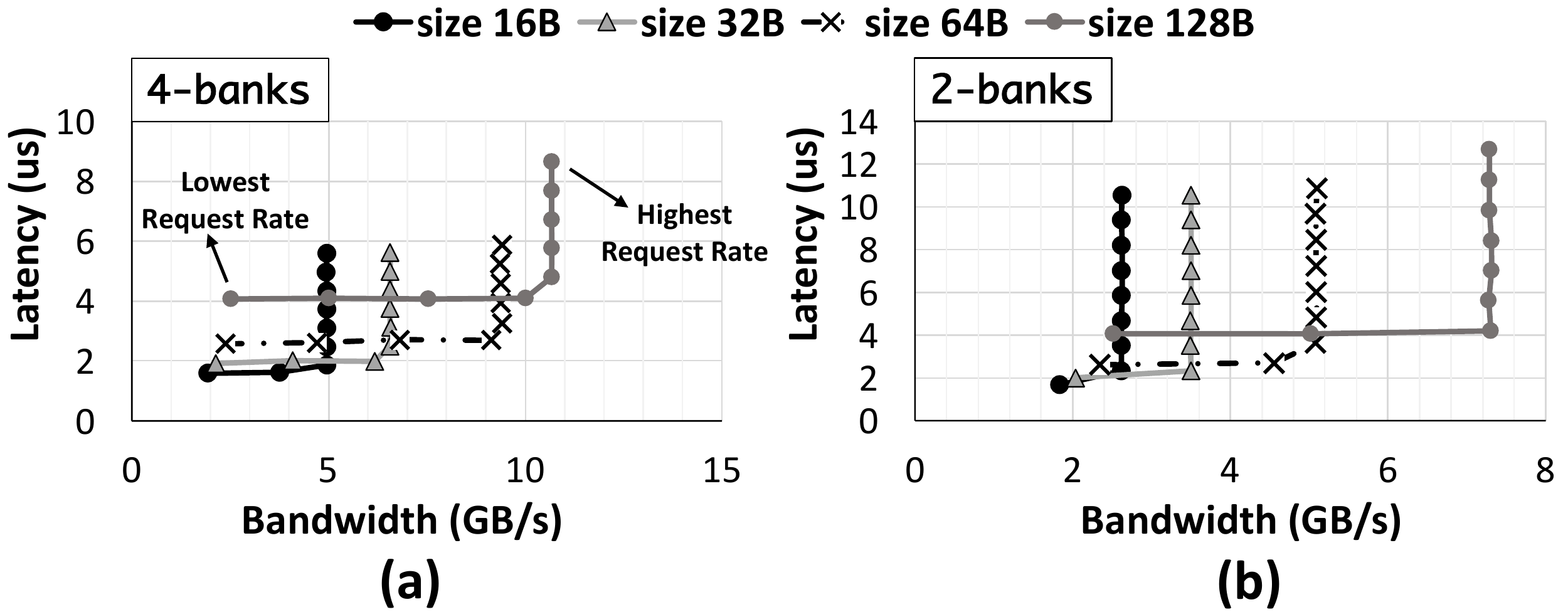}
\vspace{-0in}
\captionsetup{singlelinecheck=on,aboveskip=-8pt,belowskip=-4pt}
\caption{Latency and request bandwidth relationships for
(a) four-banks, and (b) two-banks access patterns.}
\label{fig:bw-lat-4banks}
\vspace{-0.15in}
\end{figure}

\emph{\textbf{4) Bandwidth-latency relationship:}}
To explore bandwidth and latency relationship, we first
examine two specific access patterns, 
and then all access patterns by using small-scale GUPS.
As pointed out in Section~\ref{sec:meth-firm}, we use small-scale GUPS to
tune the request rate by changing the number of active ports.
Figure~\ref{fig:bw-lat-4banks}a presents the results for read accesses
to four banks within a vault.  
The figure, an actual
demonstration of Figure~\ref{fig:moti}c, shows that latency saturates
at some rates that vary according to the packet size.  
This limiting factor that causes saturation can be the size of a queue in the vault
controller.  
By assuming that a vault controller is a black box that
includes a server and a queue, the average number of requests in a vault controller can be calculated
based on Little's law by multiplying the average time a request spends
in the controller (i.e., vertical axis) by the input rate (i.e.,
horizontal axis) at the saturated point, the result of which 
provides the number of outstanding requests in terms of bytes.  
By accounting for the request sizes,
we achieve a constant number, 375.
\begin{figure}[b]
\centering
\vspace{-0.15in}
\includegraphics[width=1\linewidth]{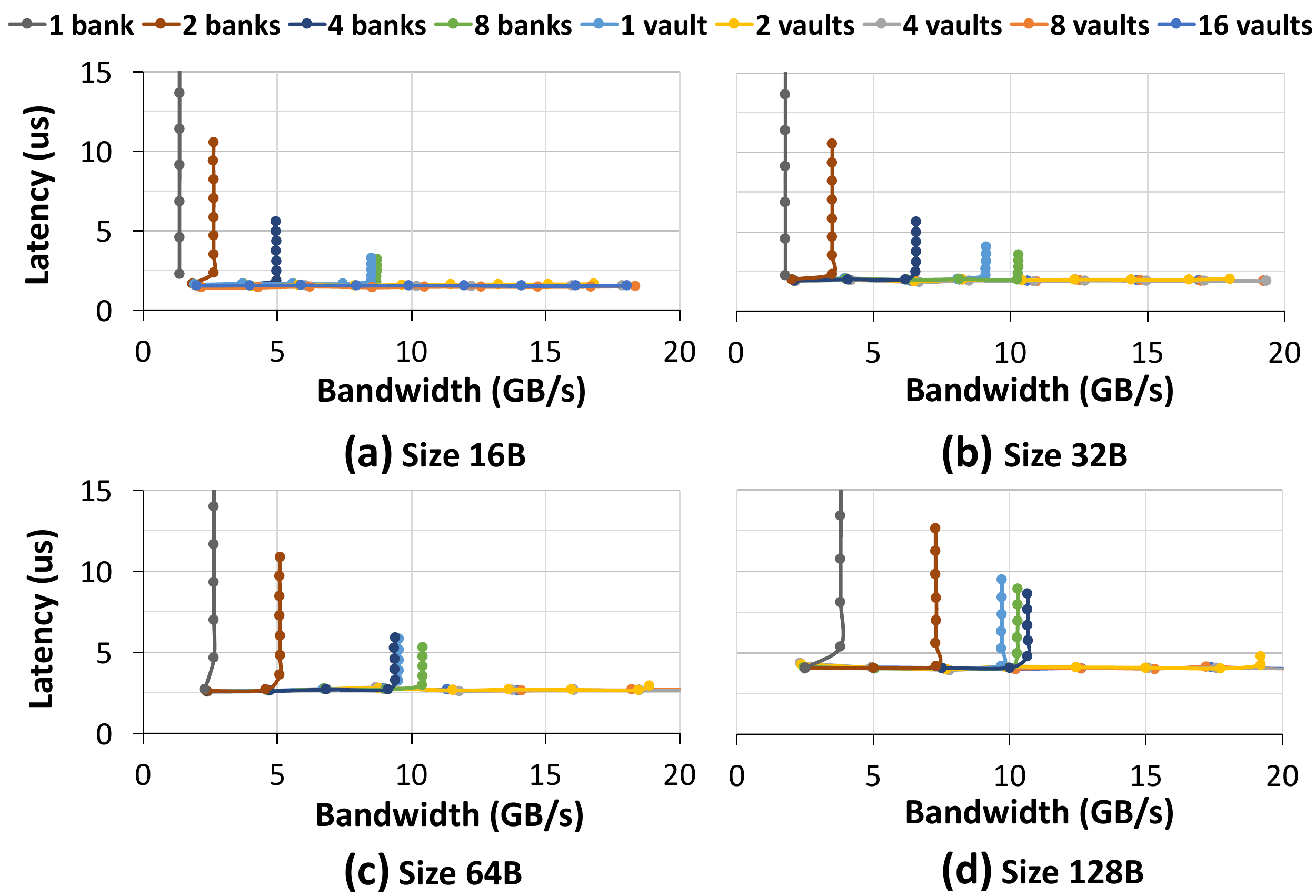}
\vspace{-0.25in}
\caption{Read latency and request bandwidth relationships for various
request sizes.}
\label{fig:bw-lat}
\vspace{-0.1in}
\end{figure}
Similar analysis of Figure~\ref{fig:bw-lat-4banks}b infers that the
number of outstanding requests for two-bank accesses is half of that for
four-bank accesses.  
This observation suggests that a vault controller
has one queue for each bank or for each DRAM layer.

To explore further the limits
of latency, Figure~\ref{fig:bw-lat} presents an extended
version of Figure~\ref{fig:bw-lat-4banks} for various sizes and access patterns.  
As shown, accesses to more than eight banks of a vault do
not follow the inferred trend of doubling the number of outstanding requests 
(i.e., benefiting from BLP) 
followed by accesses to one, two, and four
banks.  
In fact, another factor, which is 10\,GB/s access
bandwidth within a vault, is responsible for this limitation. 
In addition, when accesses are spread over two vaults, the saturation point occurs at
19\,GB/s, which is about 2x as high as that at 10\,GB/s, the bandwidth
limit of a vault.  
Notice that since we could not generate more
parallel accesses to the HMC, the saturation points of accesses to
more than two vaults does not occur.  
These results reiterate the importance of understanding the internal
structure and abstractions of HMC for maximizing application
performance. 
Besides, some limits exist on vault bandwidth (10\,GB/s) that
can be exceeded by appropriate data layout schemes. 
Further, we note the asymmetry in requests, for example each read request can consume
128\,B of return bandwidth while the request itself takes only 16\,B. 
Thus, achievable bidirectional bandwidth is derated from the peak
bandwidth. 
Appropriate data mappings are necessary to not derate
achievable bandwidth further.

\section{Related Work}
\label{related}

\noindent HMC is a practical 3D-stacked
memory that integrates  DRAM dies and computational logic. 
Understanding the features of HMC is crucial for 
state-of-the-art applications that want to benefit 
from high bandwidth or low latency.
Therefore, this work, in continuation of previous studies, 
have explored the characterization of HMC in terms of bandwidth, 
latency, thermal behavior, and power consumption. 
Recently, actual hardware-based studies continued the trend of HMC characterization, 
started by simulation-based explorations.
In a technical report, Rosenfeld et al.~\cite{ros:pau2012}
explored HMC performance in terms of bandwidth, latency and power consumption
under HMCSim simulation.
They showed that maximum link efficiency is achieved by a read ratio between 53\% to 60\%
(based on packet size).
The conclusion was while only memory-intensive applications achieve remarkable speedup
from HMC, all applications may benefit from the
low power consumption of HMC.
In the first real-world HMC characterization work, 
Gokhale et al.~\cite{gok:llo15} 
measured bandwidth and latency of HMC\,1.1 by running 
a set of data-centric benchmark using an FGPA emulator.
They showed that read-dominated applications achieve a 80\% of 
peak bandwidth. 
Moreover, they discern that high concurrency of accesses leads to 
an exponential increase in latency. 
Furthermore, they reported a correlation between latency and bandwidth, 
in which latency changes from 80\,ns for serial workloads, 
to 130\,ns for concurrent workloads, while the bandwidth 
increases from 1\,GB/s to 65\,GB/s and power consumption increases 
from 11\,W to 20\,W. 

In another real-world characterization study, Schmidt et al.~\cite{sch:fro2016} 
characterized HMC\,1.0 using OpenHMC, an open-source digital design (i.e., infrastructure).
They confirmed previously mentioned simulation-based results that
a read ratio between 53\% to 66\% maximizes the 
access bandwidth of HMC. 
Moreover, they showed that read latency increases 
dramatically to more than 4000\,ns, if the read ratio exceeds  
the optimal rate. 
However, for less than 53\% read ratio, 
a total read latency of 192\,ns and 224\,ns (for 12.5 and 10\,Gpbs respectively) is reported. 
Also, the power consumption, measured in that paper is up to 9\,W. 
In another experimental paper, 
Ibrahim et al.~\cite{ibr:fat2016} analyzed the effect of spatial and temporal 
locality of accesses to HMC on its throughput with an EX800 
backplane consisting of four Stratix FPGAs.
They emphasized that access patterns with high temporal and spatial locality
can operate up to 20 times as fast as other applications. 
Similar to our result, their outcome reveals that write bandwidth is lower 
than read bandwidth. 
While these studies, including our work, agree on the general performance behaviors 
in terms of bandwidth and latency, as a result of varied platforms, some of the provided
results are slightly different.
Although previous work have illustrated fascinating facts about the features of HMC,
these real-world characterization studies have not covered the impact of utilizing 
desirable features of HMC on temperature, power consumption, and latency. 
In the meantime, some of the recent studies explored thermal behavior of HMC by simulation.
Zhu et al.~\cite{zhu:yux:2016} analyzed the thermal behavior 
of PIM systems, in which processing elements are 
integrated with die-stacked memory.
They show that a higher ambient temperature 
lowers the power budget for each functional unit at PIM. 
Also, to provide a safe operational environment for PIM, 
a stronger cooling solution 
(i.e. costlier in terms of price and power consumption) should be used. 
In addition,
Eckert et al.~\cite{yas:nuw14} evaluated various cooling solutions for PIM-based systems. 
Their simulation using HotSpot showed that 
a low-cost passive heatsink is enough for keeping DRAM under the threshold 
temperature ($85\degree$C), 
while provisioning sufficient power for computations in memory. 
While simulation-based thermal analysis in these works yields useful insights, it does not accurately reflect the actual issues and the correlation between bandwidth, temperature, and power consumption.

\vspace{0pt}
\section{Conclusion}
\label{conclusion}

\noindent
This work demystified the characteristics of HMC, one of the two commodity 3D-DRAM technologies, which represents a class of 3D memories with high internal concurrency.
The insights and projected results are generic not only to the class of  3D-memory systems but also to architectures that utilize packet-switched communications.
Our experiments revealed the following key insights: 
(\romannum{1}) To efficiently utilize bi-directional bandwidth, accesses should have large sizes and use a mix of reads and writes.
(\romannum{2}) To avoid structural bottlenecks and to exploit BLP, accesses should be distributed and request rate should be controlled from any level of abstraction (i.e., OS, compiler, software, or hardware).
(\romannum{3}) To avoid complexity, one should not consider to improve performance resulting from spatial locality.
(\romannum{4}) To benefit from scalability of packet-switched communications, a low-latency infrastructure is crucial.
(\romannum{5}) To enable temperature-sensitive operations, fault-tolerant mechanisms should be employed.
(\romannum{6}) To attain high bandwidth, optimized low-power mechanisms should be integrated with proper cooling solutions.
While some experiments in this papers provide anticipated results, they yield a concrete real measured data to verify and also to explore future architecture research for an emerging technology.

\section*{Acknowledgments}
\noindent
We thank anonymous reviewers for their valuable comments 
and feedbacks for improving the paper. 
Our experimental hardware is partially supported by Micron.
This study was supported in part by National Science 
Foundation under grant number CCF-1533767. 

\def\IEEEbibitemsep{0pt plus 0pt}

\vspace{-5pt}
\bibliographystyle{IEEEtran}
\bibliography{short}

\end{document}